\documentclass[aps,pre,
reprint,
 notitlepage,
 showpacs,floatfix,
 nofootinbib,
 superscriptaddress
 ]{revtex4-1}

 \usepackage{subfigure}
 \usepackage{amssymb}
 \usepackage{amsfonts}
 \usepackage{amsmath}
 \usepackage{amsthm}
 \usepackage{epsfig}
 \usepackage{float}
 \usepackage[usenames,dvipsnames]{color}
 \usepackage[latin1]{inputenc}

 \usepackage{graphics,graphicx,xcolor,subfigure}

 \usepackage[hidelinks,unicode=true]{hyperref}
 \hypersetup{colorlinks=true,
 	linkcolor=blue,
 	urlcolor=blue,
 	citecolor=blue,
 	pdfhighlight=/N
 }
 
 \usepackage{comment}
 \usepackage[normalem]{ulem}
 \usepackage{microtype}

\newcommand{\av}[1]{\langle {#1} \rangle}

\newcommand{\Ninf}{N_\mathrm{inf}}

\newcommand{\NSI}{N_\mathrm{SI}}
\newcommand{\kmax}{k_\text{max}}

\newcommand{\kc}{k_\text{c}}

\begin{document}

\title{Dissecting localization phenomena of dynamical processes on networks}

\author{Diogo H. Silva}
\affiliation{Departamento de F\'{\i}sica, Universidade Federal de Vi\c{c}osa, 36570-900 Vi\c{c}osa, Minas Gerais, Brazil}


\author{Silvio C. Ferreira}
\affiliation{Departamento de F\'{\i}sica, Universidade Federal de Vi\c{c}osa, 36570-900 Vi\c{c}osa, Minas Gerais, Brazil}
\affiliation{National Institute of Science and Technology for Complex Systems, 22290-180, Rio de Janeiro, Brazil}

\begin{abstract} 
Localization phenomena permeate many branches of physics playing a fundamental role on dynamical processes evolving on heterogeneous networks. These localization analyses are frequently grounded, for example, on eigenvectors of adjacency or non-backtracking matrices which emerge in  theories of dynamic processes near to an active to inactive transition. We advance in this problem gauging nodal activity to quantify the localization in dynamical processes on networks whether they are near to a transition or not. The method is generic and applicable to theory, stochastic simulations, and real data. We investigate spreading processes on a wide spectrum of networks, both analytically and numerically, showing that nodal activity can present complex patterns depending on the network structure. Using annealed networks we show that a localized state at the transition and an endemic phase just above it are not incompatible features of a spreading process. We also report that epidemic prevalence near to the transition  is determined by the  delocalized component of the network even when the analysis of the inverse participation ratio indicates a localized activity. Also, dynamical processes with distinct critical exponents can be described by the same localization pattern. Turning to quenched networks, a more complex picture, depending on the type of activation and on the range of degree exponent,  is observed and discussed. Our work paves an important  path for investigation of localized activity in spreading and other processes on networks. 
\end{abstract}


\maketitle

\section{Introduction}
\label{sec:intro}

Statistical mechanics of phase transitions was  devised for infinitely large systems  due to their singular behavior, possible only in the thermodynamic limit~\cite{stanley1987introduction}, in which subleading contributions from finite size regions to the order parameter become irrelevant. Dynamical processes that undergo nonequilibrium phase transitions share these same properties despite lacking a general background as in the equilibrium statistical mechanics~\cite{Henkel2008}. However, the contribution of localized regions can change drastically the nature of the phase transition as, for example, its smearing which caused by extended disorder~\cite{Vojta2006} in subextensive domains where the order parameter is locally nonzero but globally represents a negligible contribution. Another interesting phenomenon is the Griffiths phase, in which critical behavior  in extended regions of the space parameter is produced by rare and locally active domains due to  randomness in disordered systems~\cite{Vojta2006}.

Localization is central for activation transitions of dynamical processes on heterogeneous networks~\cite{Goltsev2012, Buono2013,Moretti2013,Odor2014, DeArruda2017a, Cota2018, Hebert-Dufresne2019,  St-Onge2020}. Let us consider dynamical systems where the agent states are represented by vertices of a network and their interactions by the edges connecting them. Examples include epidemic processes~\cite{PastorSatorras2015}, opinion
formation~\cite{Castellano2009}, and synchronization~\cite{Rodrigues2016} models. One can define an order parameter that determines in which phase the system is. For example, a nonnull epidemic prevalence is the fraction of infected individuals  that can indicate an endemic phase where the epidemic thrives in an extensive part of the system. Prevalence is  an order parameter in the range $[0,1]$ that can be defined globally, locally, or even for a single vertex and represents an activity and can be generalized for other processes than epidemics. The local activity could be the probability that a vertex is infected in an epidemic or  has the majority opinion in a voter-like model. Subgraphs such as hubs and their neighbors~\cite{Castellano2012} or cliques of densely connected subgraphs~\cite{St-Onge2020} can hold high local activity  for very long times (metastability) even if the overall order parameter is asymptotically null.

Dynamical processes on networks are frequently investigated within mean-field approaches~\cite{PastorSatorras2015,Castellano2009,DeArruda2018}, in which the transition between zero and nonzero order parameter is the main subject. When the transition is smooth, with the order parameter approaching zero continuously, these theories can be handled perturbatively at low prevalences using linear stability analyses~\cite{PastorSatorras2015,DeArruda2018} to investigate the behavior at  the transition  in terms of the spectral properties of matrices which are related to the network structure~\cite{Goltsev2012,	Mieghem2012, Karrer2010, Mata2013, Shrestha2015, Silva2019}. A fundamental example is the susceptible-infected-susceptible (SIS) epidemic model, investigated thoroughly in the present work. In the SIS model, vertices can be infected  or susceptible. Infected vertices become susceptible with rate $\mu$  and infect its susceptible nearest-neighbors with rate $\lambda$ per contact. Let $\rho_i$ be the probability that vertex $i=1,\ldots,N$ is infected and $\rho=\sum_i \rho_i/N$ the global prevalence. The evolution can be investigated using the quenched mean-field (QMF) theory encoded in the equation~\cite{Goltsev2012,PastorSatorras2015}
\begin{equation}
\frac{d\rho_i}{dt} = -\mu\rho_i+\lambda(1-\rho_i)\sum_{j=1}^{N}A_{ij}\rho_j,
\label{eq:QMF}
\end{equation}
in which $A_{ij}$ is the adjacency matrix defined as $A_{ij}=1$ if $i$ and $j$ are connected and $A_{ij}=0$ otherwise.  The system undergoes a transition from an active to a disease-free state at the epidemic threshold 
\begin{equation}
\lambda_\text{c}^\text{QMF} =\frac{\mu}{\Lambda_1},
\end{equation}
where $\Lambda_1$ is the largest eigenvalue (LEV) corresponding to the principal eigenvector  (PVE) $\boldsymbol{v}^{(1)}=\lbrace v_1^{(1)},\ldots,v_N^{(1)}\rbrace$ of the adjacency matrix~\cite{Goltsev2012,Mieghem2012}. Moreover, we have  $\rho_i\propto
v_i^{(1)}$ and ~\cite{Goltsev2012,Mieghem2012}
\begin{equation}
\rho\simeq a_1 (\lambda/\lambda_\text{c}-1)^{\beta_\text{QMF}},
\label{eq:rhocritqmf}
\end{equation}
for $\lambda-\lambda_\text{c}\ll \lambda_\text{c}$ where $\beta_\text{QMF}=1$ and 
\begin{equation}
a_1= \frac{\sum_{i=1}^{N} v^{(1)}_i}{N \sum_{i=1}^{N} \left[v^{(1)}_i\right]^3}.
\label{eq:alpha1}
\end{equation} 
Since $\Lambda_1$ diverges as $N\rightarrow\infty$ for random networks with power-law degree distribution of the form $P(k)\sim k^{-\gamma}$ irrespective of the degree exponent $\gamma>2$~\cite{Chung2003}, one has an asymptotically null QMF threshold for SIS on these networks.

Based on Eqs.~\eqref{eq:rhocritqmf} and \eqref{eq:alpha1}, Goltsev \textit{et al}.~\cite{Goltsev2012} raised the question of whether the QMF prevalence near to the transition, when $\rho_i\propto v_i^{(1)}$, corresponds to an intensive quantity or not: An actual endemic phase demands that the PVE is delocalized and $a_1$ is nonzero whereas a localized active phase holds if $a_1\rightarrow 0$ as $N\rightarrow\infty$. Note, however, that Eq.~\eqref{eq:rhocritqmf}  is valid for $\lambda-\lambda_\text{c}\ll \lambda_\text{c}$~\cite{Goltsev2012}, representing a trick range since $\lambda_\text{c}\rightarrow 0$.  In general, for finite networks the larger the spectral gap $\Lambda_1-\Lambda_2$,  where $\Lambda_2$ is the second largest eigenvalue of $A_{ij}$, the more accurate Eq.~\eqref{eq:rhocritqmf} is~\cite{Mieghem2012}.

Goltsev et al.~\cite{Goltsev2012}  used  the inverse participation ratio (IPR) of the normalized PVE, defined as~\cite{Goltsev2012} 
\begin{equation}
Y_4(\boldsymbol{v}^{(1)})=\sum_{i=1}^{N}\left[v_i^{(1)}\right]^4,
\end{equation}
to quantify localization in the QMF theory. The IPR assumes a finite value as $N\rightarrow \infty$ for nonextensive localization on finite set of vertices, vanishes as $Y_4\sim N^{-\nu}$ with $0<\nu<1$ for subextensive localization, and as $Y_4\sim N^{-1}$ if the PVE is delocalized (extensive). According to this classification, the PVE for  random networks is localized in a nonextensive region for  degree exponent $\gamma>5/2$  and in a subextensive region for $\gamma<5/2$~\cite{Pastor-Satorras2016}, and in both cases the QMF result $\rho_i\propto v_i^{(1)}$ corresponds to a localized phase. Moreover, the exponent $\beta_\text{QMF}=1$ in Eq.~\eqref{eq:rhocritqmf} is in odds with the exact result of Mountford et al.~\cite{Mountford2013} where $\beta>1$ for any $\gamma>2$. However, recent numerical integration  of the QMF equations and its extension to a pair approximation~\cite{Silva2020} show that these theories can fit simulations very accurately if not too close to the epidemic threshold, which is out of validity range of Eq.~\eqref{eq:rhocritqmf}.

A message-passing theory for susceptible-infected-recovered (SIR) model, for which an infected individual becomes recovered and cannot be reinfected (long-term immunization)~\cite{PastorSatorras2015}, provides an epidemic threshold  given by the inverse of LEV of the non-backtracking or Hashimoto matrix~\cite{Hamilton2014} constructed with the set of edges of the networks; see Refs.~\cite{Hamilton2014,Shrestha2015,Pastor-Satorras2020} for more details on Hashimoto matrices. In addition to the well-documented PVE localization of the adjacency matrix and its implication to the SIS dynamics, it was recently discussed localization in the PVE associated to the non-backtracking matrix on real networks and its implications for  SIR and percolation processes~\cite{Pastor-Satorras2020}.

The aforementioned results point out that localization analyses based on the spectral properties of particular matrices are limited and can lead to misleading conclusions with respect to the localization properties of the dynamical processes and the mean-field theories associated to them. Aspects  such as dynamical correlations, stochastic fluctuations, and nonperturbative contributions cannot be fully reckoned in these spectral approaches. In the present work, we investigate localization of epidemic activity considering the actual dynamics on networks using a normalized activity vector (NAV) that gauges the contribution of each vertex of the network to the total activity. We analyze the SIS~\cite{PastorSatorras2015} and contact process (CP)~\cite{Mata2014} models due to their very distinct mechanisms of activation~\cite{Ferreira2016a}, but the methodology is not limited to them. Using both analytical and simulation tools, we show that the NAV components are directly associated to the vertex degree in annealed networks. Localization patterns for critical CP in annealed and quenched networks are essentially the same, corroborating that the critical behavior of this dynamical system is described by a  heterogeneous mean-field (HMF) theory~\cite{Ferreira2011,Mata2014}. However, localization  of critical SIS on quenched networks partially agrees with annealed networks for $\gamma<5/2$ whereas highly structured and complex localization patterns emerge on quenched networks with $\gamma>5/2$. We also show that CP and SIS on annealed networks present localization at the transition despite the delocalized, active phase that emerges just above it. Using fractional averages we show that the order parameter near to the transition is ruled by contributions of lowly localized vertices, which are by far the vast majority, while the IPR is determined by a vanishing fraction of the network. Finally, we analyzed the accuracy of the QMF theory in predicting the localization patterns observed in stochastic simulations of the SIS near to the epidemic threshold, and found that it is very accurate for $\gamma<5/2$ but deviates quantitatively otherwise.

The remainder of the paper is organized as follows. Definition of the NAV, its determination, models, and other methods used in this work are presented in Sec.~\ref{sec:models}. Effects of localization and their interpretation are exemplified applying  the NAV tool to two simple networks models in Sec.~\ref{sec:simple}. The central core of the work, the analysis of localization on random networks with power-law degree distributions, is presented in Sec.~\ref{sec:locPL}. We finally draw our concluding remarks in Sec.~\ref{sec:conclu}.

\section{Models and  methods}
\label{sec:models}

Localization of dynamical processes on networks can be quantified using a local order parameter that measures the level of activity in a determined subset. We consider processes on the top of networks composed of $i=1,2,\ldots,N$ vertices or nodes and a set of edges or links that represent the interactions among them.  The state of each vertex is represented by $\sigma_i$ and can be in one of two classes: active or inactive with probabilities $\rho_i$ and $1-\rho_i$, respectively.  We assume that the active vertices define the order parameter globally in the steady state ($t\rightarrow\infty$) as 
\begin{equation}
\rho =\frac{1}{N}\sum_{i=1}^{N}\rho_i,
\end{equation}
which gives the stationary epidemic prevalence in both SIS and CP. The concept of active vertices is trivial in epidemic models with recurrent states such as the SIS. For the SIR model, the activity can be defined as the probability that a vertex will be infected in an outbreak started in a randomly selected vertex. For synchronization or opinion models the activity will depend explicitly on the neighborhood and should take into account higher order interactions such as pairwise. In this work, we deal only with the cases of  activity defined at the nodal level since the generalization to higher order motifs  can be done accordingly.

We define the activity vector $\boldsymbol{\rho}=(\rho_1,\ldots,\rho_N)$. We investigate processes with absorbing states on finite networks. We overcome the difficulties inherent to these systems by introducing a small self-activation~\cite{VanMieghem2020}  of vertices. It can be an exogenous infection in epidemics or, in general, spontaneous creation in reaction-diffusion processes. If the process has active steady states, self-activation  can be introduced through a uniform and spontaneous rate $f$ such that $f\rightarrow 0$ as $N\rightarrow\infty$~\cite{Sander2016}. One advantage of this approach is that localization can be investigated also in the absorbing phase of the original dynamics. Obviously, other approaches, such as quasi-stationary simulations~\cite{Sander2016}, can be used too.

In analogy to the PVE, we  introduce the NAV $\boldsymbol{\phi}=(\phi_i,\ldots,\phi_N)$ whose components are given by
\begin{equation}
\phi_i=\frac{\rho_i}{\sqrt{\sum_{j=1}^{N} \rho_j^2}}.
\label{eq:phii}
\end{equation}
Two limit cases are the completely localized and delocalized states where $\rho_i=N\delta_{ij}\bar{\rho}$ and  $\rho_i=\bar{\rho}$, respectively. The corresponding  NAV components are $\phi_i=\delta_{ij}$ and
$\phi_i=1/\sqrt{N}$, respectively. The IPR is defined as
\begin{equation}
Y_4(\boldsymbol{\phi})=\sum_{i=1}^{N} \phi_i^4,
\end{equation}
for which we derive the limit behaviors of $Y_4=N^{-1}$ for fully delocalized and $Y_4=1$ for fully localized activity.

We analyzed localization in two recurrent activation models with binary states, namely, the SIS model~\cite{PastorSatorras2015} and the CP~\cite{Henkel2008}. Contact process is a variation of the SIS model, in which the infection rate per contact is $\lambda/k_i$ where $k_i$ is the degree of the vertex  which is transmitting the infection~\cite{Mata2014} while the healing process is exactly the same of the SIS. In both cases, states of a vertex can be represented by $\sigma_i=0$ when the vertex is inactive (susceptible) or $\sigma_i=1$ when it is active (infected). Despite similarities between the models, their activation processes 
are of very different nature when evolving on random power-law networks~\cite{Ferreira2016a}. In the SIS, activation is triggered by hubs while in CP it happens collective as in an usual phase transition~\cite{stanley1987introduction}. In the SIS on random networks with $P(k)\sim k^{-\gamma}$, hubs responsible for the activation can belong to a densely connected core for $\gamma<5/2$  or be sparsely distributed for $\gamma>5/2$~\cite{Castellano2012,Pastor-Satorras2016,Ferreira2016a}. Thus, different patterns of localization are eligible for these models.

We investigated the models using  HMF~\cite{Pastor-Satorras2001}  and QMF~\cite{Goltsev2012} theories as well as stochastic simulations~\cite{Cota2017}. The stochastic simulations of the SIS with a source were performed using the following optimized Gillespie algorithm based on phantom
processes~\cite{Cota2017}. At each time step,  with probability 
\begin{equation}
P=\frac{\mu\Ninf}{\lambda \NSI+fN+\mu\Ninf}
\label{eq:Pheal}
\end{equation}
one infected vertex is chosen  at random and becomes susceptible. Here $\Ninf$ is the total number of infected vertices and $\NSI$ is the total number of links emanating from them.  With probability
\begin{equation}
Q=\frac{fN}{\lambda \NSI+fN+\mu \Ninf},
\label{eq:Pinf}
\end{equation} 
a vertex of the network is chosen with equal chance and becomes infected if it is susceptible. Finally, with probability $1-Q-P$ an infected vertex $i$ is selected with probability proportional to its degree $k_i$ and one of its neighbors $j$ is chosen with equal chance. If $j$ is susceptible it becomes infected. The time is incremented by
\begin{equation}
\delta t=\frac{-\ln \xi}{\lambda \NSI+fN+\mu\Ninf},
\label{eq:deltat}
\end{equation}
where $\xi$ is a random number uniformly distributed in the interval $(0,1)$. In CP simulations we replace $\NSI$ by $\Ninf$ in Eqs.~\eqref{eq:Pheal}, \eqref{eq:Pinf}, and \eqref{eq:deltat} and chose  with equal chance the infected vertex  that transmits while all other steps are identical. See Ref.~\cite{Cota2017} for more details on phantom processes.
We compute $\boldsymbol{\rho}$ in simulations as the fraction of time that each vertex $i$ is active during an averaging time of $t_\text{av}$  after a relaxation time of $t_\text{rlx}$. Mathematically it is expressed as
\begin{equation}
\boldsymbol{\rho} = \frac{1}{t_\text{av}}\int_{t_\text{rlx}}^{t_\text{rlx}+t_\text{av}} \boldsymbol{\sigma}(t)dt.
\end{equation} 
Values  $t_\text{av}=10^5~\mu^{-1}$ to $10^8~\mu^{-1}$ and $t_\text{rlx}=10^4~\mu^{-1}$ to $10^6~\mu^{-1}$ were considered, being the larger times used for lower activity regimes where fluctuations are larger. In numerical integrations, $\boldsymbol{\rho}$ is obtained after convergence to the steady-state. In all simulations we used $f=\mu/N$ meaning that at most one new infection is randomly introduced by one unit of time $1/\mu$.

\section{Localization phenomena  on simple networks}
\label{sec:simple}

In order to obtain insights about the localization in terms of the NAV, we start our analyses with simple networks

\subsection{Star graph}
\label{subsec:star}

A star graph consists of a central node $i=0$ connected to $i=1,\ldots,K$ neighbors of degree $k=1$ (leaves). The adjacency matrix is $A_{0i}=A_{i0}=1$ for $i>0$ and $A_{ij}=0$ otherwise.  Let us start with the QMF theory for SIS on the star graph defining $\rho_0$ as the probability that the center is infected and $\rho_1=\rho_2=\ldots=\rho_N$ as the probability that a leaf is infected. The QMF equations become
\begin{eqnarray}
\frac{d\rho_{0}}{dt}&=&-\mu\rho_{0}+\lambda K (1-\rho_{0})\rho_{1}+f(1-\rho_{0}) \label{eq:star0}\\
\frac{d\rho_{1}}{dt}&=&-\mu\rho_{1}+\lambda (1-\rho_{1})\rho_{0}+f(1-\rho_{1}).
\label{eq:star1}	
\end{eqnarray}
Without self-activation ($f=0$) the epidemic threshold is $\lambda_\text{c}={\mu}/{\sqrt{K}}$~\cite{Ferreira2012}. The steady-state solution for $f\ll \mu$ and $\lambda\ll \mu$ is given by
\begin{equation}
\rho_{0}=\frac{\lambda K\rho_{1}+f}{\lambda K\rho_{1}+\mu}
\end{equation}
and
\begin{equation}
2\rho_{1}={\frac{\lambda+f}{\mu}-\frac{\mu}{\lambda K}+\sqrt{\left(\frac{\lambda+f}{\mu}-\frac{\mu}{\lambda K}\right)^2+\frac{4f}{\lambda K}}}.
\end{equation}
At $\lambda=\lambda_c=\mu/\sqrt{K}$ we have $\rho_0=\sqrt{f/\lambda}$ and $\rho_1=\sqrt{\lambda f}/\mu$ leading to an IPR given by  $Y_4=\phi_0^4+K\phi_1^4$, which asymptotically assumes $Y_4=1/4$ confirming localization. For supercritical region $\lambda^2 K\gg \mu$ we have $\rho_0\approx 1$ and $\rho_1={\lambda}/{\mu}$  implying $Y_4=1/K$ in a fully delocalized state. Finally, for subcritical region with $\lambda^2 K\ll \mu^2$ but still $\lambda K\gg \mu$ and $fK\sim \mathcal{O}(1)$ we obtain $\rho_0=f\lambda K/\mu^2$ and $\rho_1=f/\mu$  leading to $Y_4=(\lambda^2 K/\mu^2)^2\ll 1$  in a subextensive localization since it goes to zero slower than $K^{-1}$.
\begin{figure}[th]
	\centering
	\includegraphics[width=0.96\linewidth]{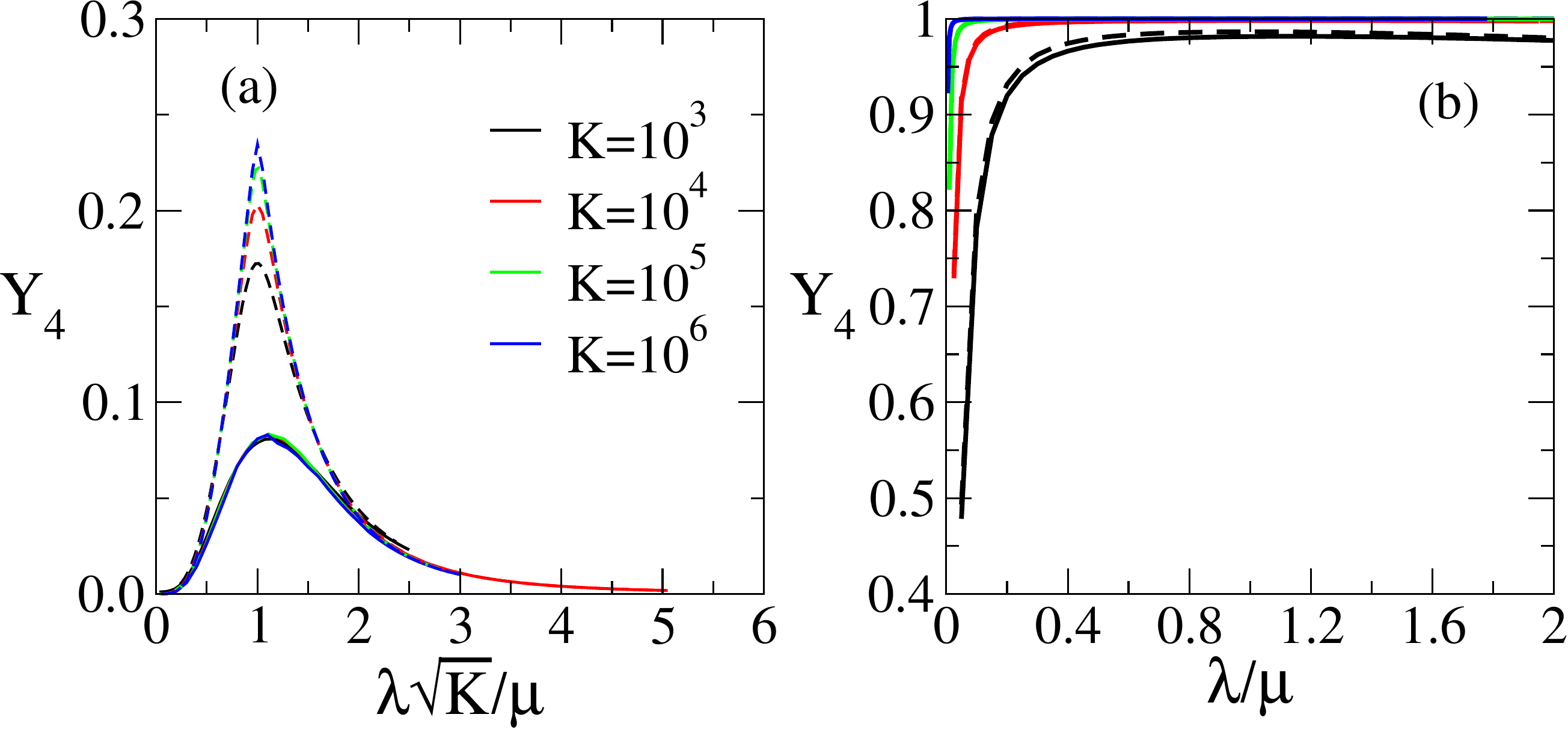}
	\caption{Inverse participation ratio analysis for (a) SIS and (b) CP models on star graphs with different number of leaves $K$. Dashed lines represent the QMF theory solution and solid lines the stochastic simulations. Dashed lines overlap with solid ones in (b) and can not be seen. }
	\label{fig:iprxlambda_star_siscp}
\end{figure}

Figure~\ref{fig:iprxlambda_star_siscp}(a) compares the IPR as a function of $\lambda\sqrt{K}/\mu$ for the QMF theory of stochastic simulations of the SIS model on star graphs of different sizes $K$. The maximum localization in the QMF theory occurs at $\lambda\sqrt{K}/\mu=1$ and the convergence to the asymptotic value $Y_4=1/4$ is verified. For simulations, the maximum value is $Y_4\approx 0.08$ occurring at $\lambda\sqrt{K}/\mu\approx 1.1$ below the SIS threshold $\lambda_\text{c}\sqrt{K}/\mu=\sqrt{2}$ for a pair QMF theory without self-activation~\cite{Mata2013}. In both cases,  the localization drops very quickly  after the transition since the range of $\lambda$ where $\lambda\sqrt{K}/\mu$ is finite corresponds to $\lambda\ll 1$.

The CP dynamics  on star graphs has a finite lifespan for $K\rightarrow\infty$ for any finite value of $\lambda\mu$~\cite{Ferreira2016a} implying that the dynamics without a self-activation is always in the absorbing phase at long times.  Developing a QMF theory for CP with self-activation, we have that $\rho_0=\lambda C(\mu,\lambda,fN)$ and $\rho_1=C(\mu,\lambda,fN)/N$ where
\begin{equation}
C(\mu,\lambda,\zeta) = \frac{\lambda^2-\mu^2+\lambda \zeta +\sqrt{(\lambda^2-\mu^2+
		\lambda\zeta)^2+\lambda \zeta \mu^2}}{2\zeta\lambda}
\end{equation}
is a finite positive constant when the arguments $(\mu,\lambda,\zeta)$ are finite. Therefore, the IPR is asymptotically $Y_4=1$, as confirmed in Fig.~\ref{fig:iprxlambda_star_siscp}(b)  for both theory and simulations, showing that the CP dynamics with self-activation on a star is fully localized in the center for any finite value of $(\mu,\lambda)$.

\subsection{Random regular networks with one outlier}
\label{subsec:rrn}

Random regular networks~(RRN) are  simple graphs where all vertices have the same degree $k_i=m$ and connections are performed at random~\cite{Ferreira2012}. They are homogeneous networks where disorder is negligible. In this network, SIS and CP become equivalent  by scaling the infection rate as $\lambda^\text{SIS}=m\lambda^\text{CP}$ and we consider only the former model. The transition for $f=0$ happens at a finite threshold, slightly above the pair approximation given by $\lambda_\text{c}^\text{SIS}=1/(m-1)$~\cite{Mata2013}, and has a homogeneous mean-field like transition~\cite{Henkel2008,Ferreira2013}. The SIS dynamics on a RRN is fully delocalized.

\begin{figure}[th]
	\centering
	\includegraphics[width=0.9\linewidth]{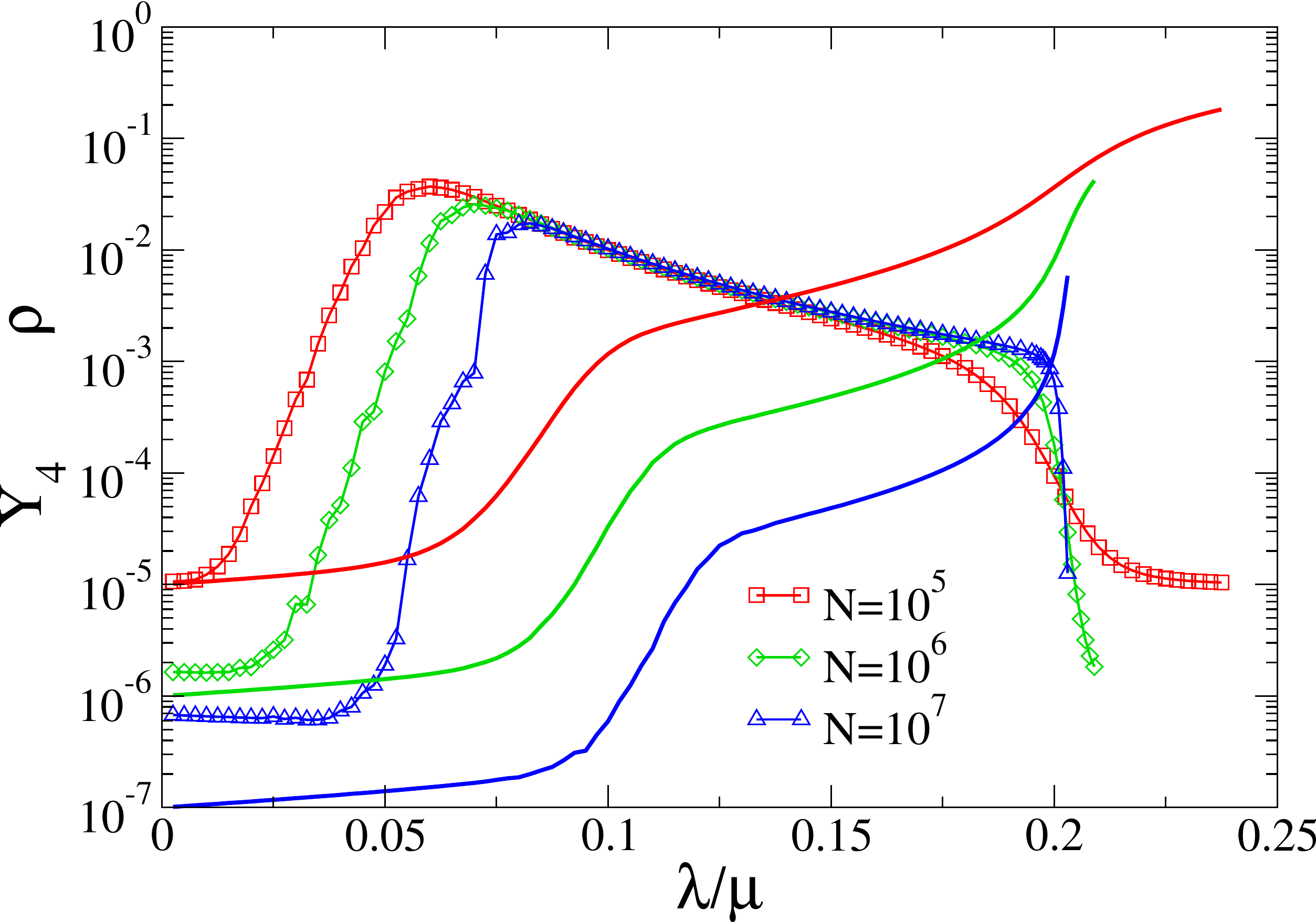}
	\caption{Inverse participation ratio $Y_4$ (lines with symbols) and prevalence $\rho$ (lines) as function of the infection rate for $f=\mu/N$. Network size is indicated in the legends. Degree of vertices is $m=6$ except one that has degree $K=10^3$.}
	\label{fig:iprxlambda_RRNstar_sis}
\end{figure}

\begin{figure}[hbt]
	\centering
	\includegraphics[width=0.988\linewidth]{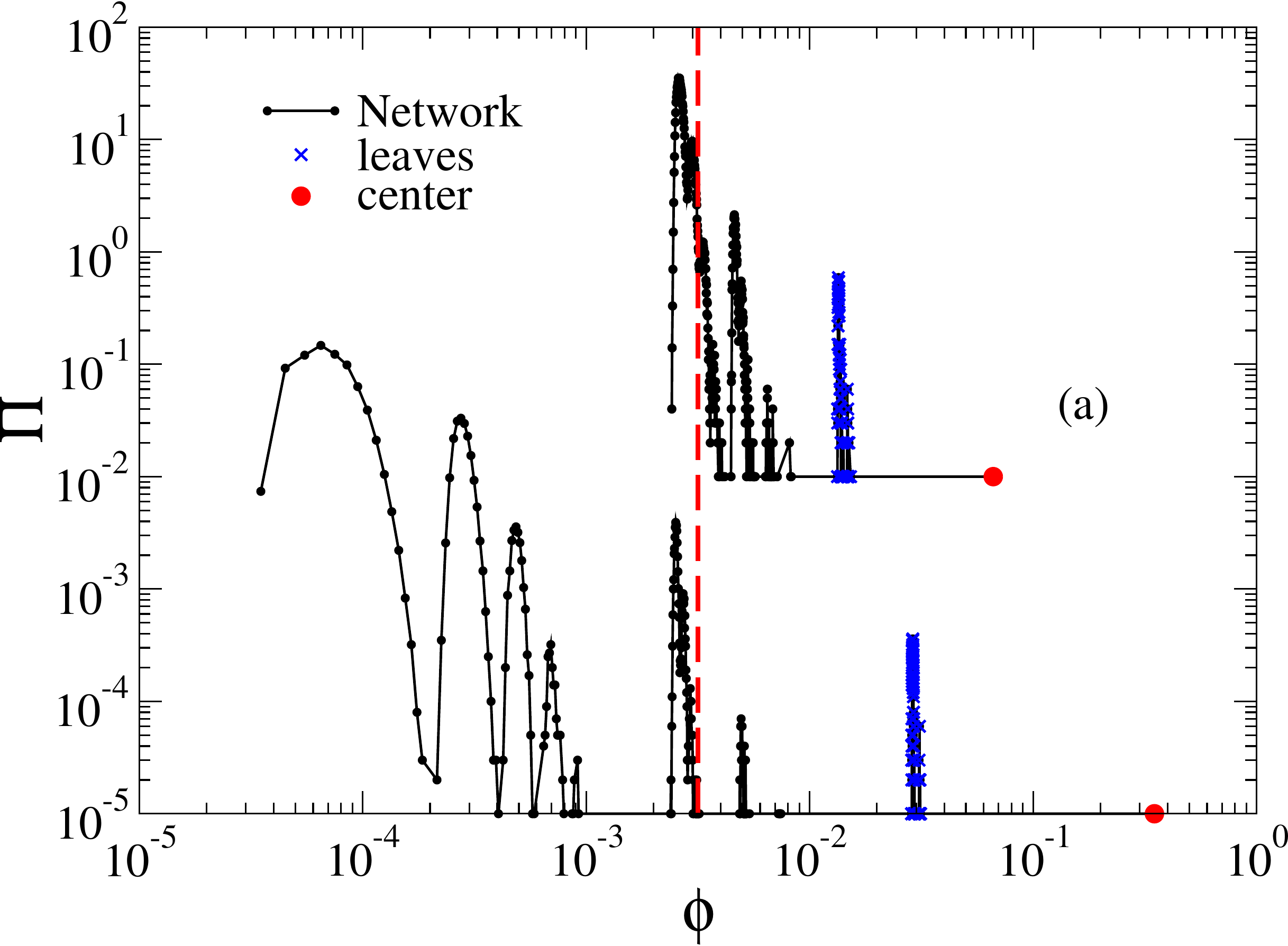}\\
	\includegraphics[width=0.988\linewidth]{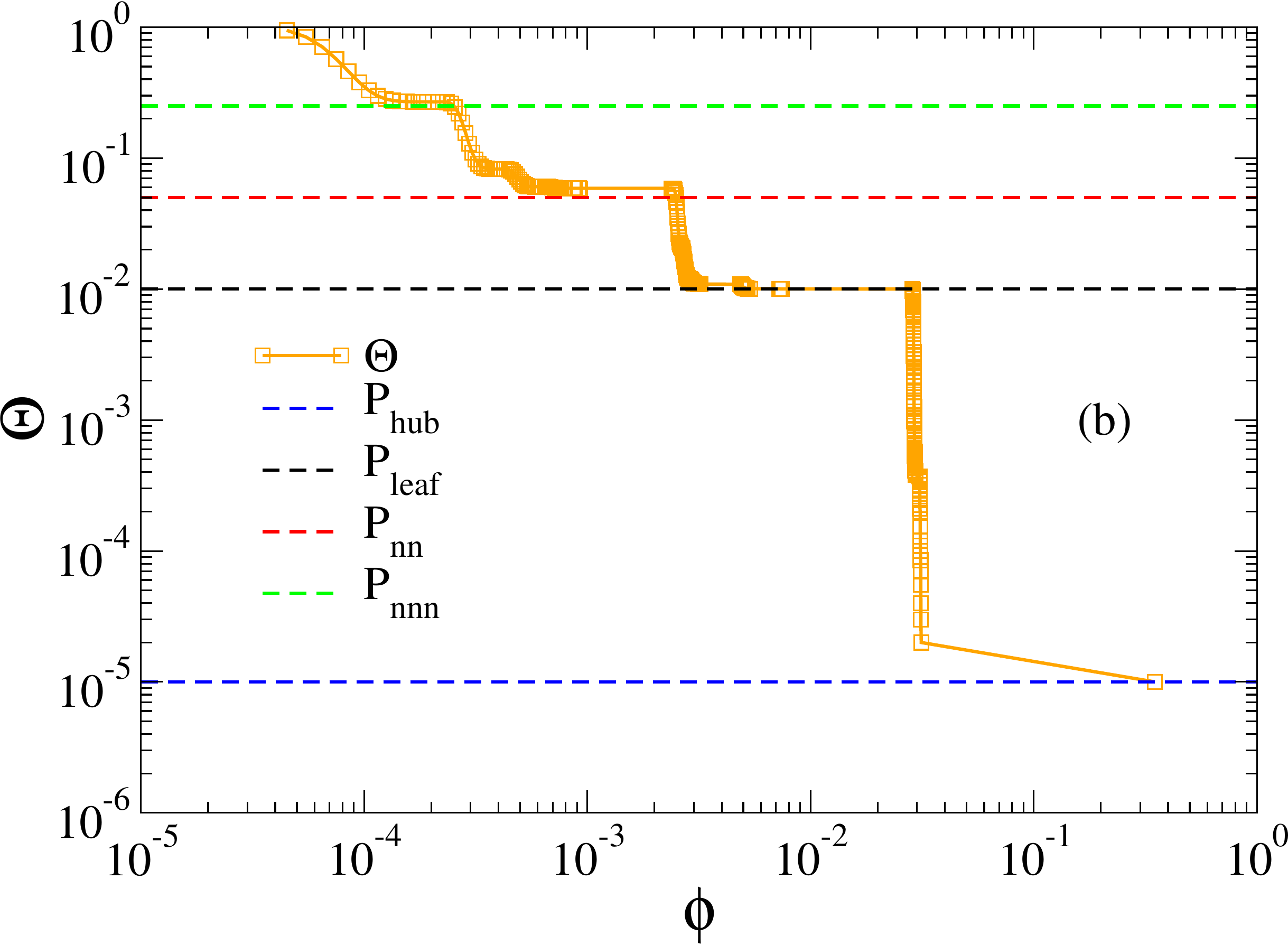}
	\caption{Analyses of the NAVCPD  for SIS on a RRN of degree $m=6$ plus a hub of degree $K=10^3$. The network size is $N=10^5$ and the binning parameter used to compute the distribution is $\delta\phi=1/N$. (a) Probability distributions for $\lambda/\mu = 0.0875$ (bottom curve), which is near to the activation of the star subgraph, and for $\lambda/\mu=0.2025$ (top curve), which near to the activation of the RRN component  and multiplied by a factor $10^3$ for sake of visibility, are shown. The components of the center and  leaves of the star subgraph are also indicated. Dashed line represents $\phi=1/\sqrt{N}$. b) Complementary cumulative probability for $\lambda=0.0875$. Dashed lines are probabilities that a randomly selected vertex belongs  to the following subgraphs: hub, leaves, nearest and next-nearest neighbors of the leaves from the bottom to the top, respectively.}
	\label{fig:dist_phi_i_SISRRNhub}
\end{figure}
Localization can be introduced by the inclusion of a single vertex of degree $K\gg m$ to form a star subgraph immersed in the RRN~\cite{Ronan2016}. The system can, in principle, present two activations being the first  one of the star subgraph and other one of the remainder of the network at $\lambda/\mu\approx1/(m-1)$. Figure~\ref{fig:iprxlambda_RRNstar_sis} shows the epidemic prevalence and IPR as functions of the infection rate for a RRN with degree $m=6$ plus a hub of degree $K=10^3$. We see two steep shoulders indicating the activations of the star subgraph and the RRN component. Remember that the transition is smoothed by the self-activation rate and would be singular only in the thermodynamic limit. Also observe the lower bound $\rho=1/N$  before the star activation. The IPR increases rapidly, reaching a maximum slightly before the first shoulder, and then decays  slowly up to $\lambda=\lambda_\text{c}^\text{RRN}$, after which drops suddenly to a value $Y_4\sim 1/N$. Therefore, one can identify an extended localized phase in the range $\lambda_\text{c}^{\star}< \lambda<\lambda_\text{c}^\text{RRN}$, ending with a sudden delocalization. A very similar system, a Bethe lattice with a hub, was investigated in the framework of QMF theory~\cite{Goltsev2012}, in which it was shown that the PEV is localized while the eigenvector of the second LEV is delocalized.

While insightful and providing  the localization intensity, the IPR does not indicate where and how localization takes place. We can extract more information considering the NAV component probability distribution (NAVCPD) defined as the fraction of vertices $\Pi(\phi)$ whose components are in the interval $\phi_i\in [\phi,\phi+\delta \phi]$. The NAVCPD for SIS dynamics near to the activation of the star subgraph is shown  in Fig.~\ref{fig:dist_phi_i_SISRRNhub}(a). The distribution shows high localization  in the hub ($\phi\approx0.35$) and leaves connected to it ($\phi\approx0.029$). However, the distribution also presents other spaced peaks representing, in principle, intermediary localization in successive shells of neighbors centered on the hub. We tested this hypothesis calculating the complementary cumulative probability (tail distribution) $\Theta(\phi)$ defined as the probability that a vertex selected at random has NAV component larger or equal than $\phi$. Figure~\ref{fig:dist_phi_i_SISRRNhub}(b) shows $\Theta(\phi)$ corresponding to the bottom distribution of Fig.~\ref{fig:dist_phi_i_SISRRNhub}(a). We can clearly see  the plateaus corresponding to the probability that a randomly selected vertex is a leaf, $P_\text{leaf}=K/N$, their nearest, $P_\text{nn}\approx (m-1)K/N$, and next-nearest, $P_\text{nnn}\approx (m-1)^2K/N$, neighbors. Observe that each plateau in $\Theta(\phi)$ corresponds to a peak in $\Pi(\phi)$. Interestingly, other localizations are also detected in the NAVCPD, as indicated by other  peaks and plateaus not associated with the aforementioned subgraphs. Finally, near to the
activation of the RRN subgraph, top  curve of Fig.~\ref{fig:dist_phi_i_SISRRNhub}(a), the NAV is concentrated around the delocalized value $\phi=1/\sqrt{N}$, but the distribution still presents localization patterns especially in leaves and center.

\section{Localization  on power-law networks}
\label{sec:locPL}

We now consider synthetic uncorrelated networks with power-law degree
distributions in the form $P(k)\propto k^{-\gamma}$ with $k=k_{0},\ldots,\kc$. 
We consider both annealed and quenched networks with fixed minimal degree
$k_0=3$. In the former, connections are rewired at random with a rate much
higher than the typical rates of the  dynamical processes evolving on them such
that dynamical correlations are completely suppressed whereas in the latter
edges are frozen and dynamical correlations are
relevant~\cite{PastorSatorras2015}.

\subsection{Annealed networks}
\label{subsec:ann}

The classical HMF theory~\cite{Pastor-Satorras2001} becomes an exact prescription of annealed networks if fluctuations are negligible. The HMF equation for the probability $\rho_k$ that a vertex of degree $k$ is infected is given by
\begin{equation}
\frac{d\rho_k}{dt} = -\mu\rho_k+\lambda k (1-\rho_k)\Omega_k+f(1-\rho_k),
\end{equation}
where $\Omega_k$ is the probability that a randomly chosen neighbor is infected. We have $\Omega_k=\sum_{k'}P(k'|k)\rho_{k'}$ for SIS~\cite{Pastor-Satorras2001} and  $\Omega_k=\sum_{k'}P(k'|k)\rho_{k'}/k'$ for CP~\cite{Castellano2006}, in which $P(k'|k)$ is the probability that a vertex of degree $k$ is connected to a vertex of $k'$. The steady-state  solution is
\begin{equation}
\rho_k=\frac{\lambda k \Omega_k+f}{\mu+\lambda k \Omega_k+f}.
\label{eq:rhokhmf}
\end{equation}
For uncorrelated networks we have $P(k'|k)=k'P(k')/\av{k}$~\cite{Boguna2004a} and $\Omega_k=\Omega$  independent of the degree. The epidemic threshold for SIS with $f=0$ is given by $\lambda_\text{c}/\mu=\av{k}/\av{k^2}$ and goes to zero  as size diverges only for $\gamma<3$~\cite{PastorSatorras2015} while for CP we have $\lambda_\text{c}=\mu$~\cite{Castellano2008}. At  $\lambda=\lambda_\text{c}$ we have $\lambda \Omega\approx \rho/\av{k} = \sum_kP(k)\rho_k/\av{k}$ in both SIS and CP models.  This result can be obtained handling Eq.~\eqref{eq:rhokhmf} in terms of hypergeometric functions and can be found in Ref.~\cite{Pastor-Satorras2001} for SIS and Ref.~\cite{Castellano2008} for CP, for example. The NAV in terms of $\rho_k$ becomes
\begin{equation}
\phi_i=\frac{\rho_{k_i}}{\sqrt{N\av{\rho_k^2}}},
\label{eq:rhoiann}
\end{equation}
while the corresponding IPR is
\begin{equation}
Y_4=\frac{\av{\rho_k^4}}{N\av{\rho_k^2}^2}.
\label{eq:Y4rhok}
\end{equation}

The localization for $\lambda=\lambda_\text{c}$ is correlated with the vertex degree in HMF theory according to Eqs.~\eqref{eq:rhokhmf} and \eqref{eq:rhoiann}: The higher the degree the more localized its activity is. According to the denominator of Eq.~\eqref{eq:rhokhmf}, the asymptotic behavior depends on $\kmax\rho$ in the regime $f\ll \rho$ which we are considering. If $\kmax\rho\ll1$ the NAV becomes
\begin{equation}
\phi_i \simeq \frac{k_i}{\sqrt{N\av{k^2}}},
\label{eq:phiann}
\end{equation}
which leads to an IPR
\begin{equation}
Y_4=\frac{\av{k^4}}{N\av{k^2}^2}.
\label{eq:Y4ann}
\end{equation}
Equations~\eqref{eq:phiann} and \eqref{eq:Y4ann} correspond to the PVE of the adjacency matrix of an annealed network given by $A_{ij}=k_ik_j/(\av{k}N)$ presented, for example, in Refs.~\cite{Goltsev2012,Pastor-Satorras2016}. For example, the critical prevalence of the CP in the quasi-stationary (QS) state, where the dynamics  returns to a previously visited configuration when the absorbing state is reached~\cite{Sander2016}, goes to zero as $\rho_\text{qs}\sim (gN)^{-1/2}$~\cite{Ferreira2011a} where $g=\av{k^2}/\av{k}^2$. We numerically checked that  the same  result holds for CP and SIS on annealed networks using self-activation $f=\mu/N$ at the effective transition point given by the maximal of  the dynamical susceptibility $\chi=N(\av{\rho^2}- \av{\rho}^2)/\av{\rho}$~\cite{Ferreira2012}.  For power-law degree distributions, we have  $\av{k^n} \sim \kmax^{n+1-\gamma}$ for $\gamma<n+1$ and $\av{k^n}\sim\text{const.}$, otherwise, implying that $\rho\kmax\sim N^{(\gamma-1-\omega)/2\omega}$ for $\gamma<3$ and $\rho\kmax\sim N^{(2-\omega)/2\omega}$ if $\gamma>3$, in which it was assumed a general scaling for the maximal degree $\kmax\sim N^{1/\omega}$ with $\omega>1$. We analyze natural $\omega=\gamma-1$ and structural $\omega=\max(2,\gamma-1)$ cutoffs~\cite{Boguna2004a,Dorogovtsev2008}. The first one emerges when $\kc=N$ as a result of finite-size realizations of a power-law distribution while the second allows to generate uncorrelated networks for any $\gamma>2$~\cite{Catanzaro2005}.

Assuming $\rho\sim(gN)^{-1/2}$ for a structural  cutoff, we obtain $\kmax\rho\ll1$ for  $\gamma>2$ leading to
\begin{equation}
Y_4\sim \left\lbrace \begin{array}{lll}
N^{-(\omega+1-\gamma)/\omega}&  & 2<\gamma<3  \\ 
N^{-(\gamma+\omega-5)/\omega}&  & 3<\gamma<5  \\ 
N^{-1}&  & \gamma>5 
\end{array}  \right.,
\label{eq:IPRstru}
\end{equation}
implying that IPR vanishes for all values of $\gamma>2$. Strong corrections to the pure power-law scaling are present for $\gamma$ near to 3 and 5. Strict delocalization, with $Y_4\sim N^{-1}$, is found only for $\gamma>5$ whereas subextensive localization, with $Y_4\sim N^{-\nu}$, and $\nu<1$~\cite{Pastor-Satorras2016}, is found otherwise. Interestingly, SIS and CP models have the same localization structure near to the transition whereas their critical behaviors for  infinite networks with $N=\infty$ and $\kc=\infty$ present different exponents for scale-free networks~\cite{Pastor-Satorras2001,Castellano2008}.

For the natural cutoff $\omega=\gamma-1$ and $\gamma<3$, assuming again $\rho\sim (gN)^{-1/2}$, we have $\rho\kmax\sim \mathcal{O}(1)$ and 
\begin{eqnarray}
\av{\rho_k^n}  & \simeq  & \int_{k_0}^{\kmax}\rho_k^n P(k) dk  =  F\left(n,\gamma-1,\gamma;-\dfrac{\av{k}}{\rho k_0}\right) \nonumber\\
&  &
-\left(\dfrac{k_0}{\kmax}\right)^{\gamma-1} F\left(n,\gamma-1,\gamma;-\dfrac{\av{k}}{\rho \kmax}\right),
\end{eqnarray}
in which $F(a,b,c;x)$ is the Gauss hypergeometric function and $n=2$ or 4. Taking the asymptotic limit  with $\rho k_0\ll 1$ and keeping only the leading contribution in the first term we obtain
\begin{eqnarray}
\av{\rho_k^n} & \simeq & \dfrac{\Gamma(n+1-\gamma)\Gamma(\gamma)}{\Gamma(n)}\left(\dfrac{\rho k_0}{\av{k}}\right)^{\gamma-1}\nonumber\\
& & 
-\left(\dfrac{k_0}{\kmax}\right)^{\gamma-1} F\left(n,\gamma-1,\gamma;-\dfrac{\av{k}}{\rho \kmax}\right),
\label{eq:rhokn1}
\end{eqnarray}
in which $\Gamma(x)$ is the Gamma function. Since $\rho\kmax\sim \mathcal{O}(1)$, so does $ F(n,\gamma-1,\gamma;-\tfrac{\av{k}}{\rho \kmax})$,  implying that both terms in Eq.~\eqref{eq:rhokn1} are of the same order $\rho^{\gamma-1}$ due to $\rho\sim1/\kmax$ and
\begin{equation}
\av{\rho_k^n} \simeq  c_n \rho^{\gamma-1}\sim 1/N,
\end{equation} 
in which $c_n$ is a constant that depends on both epidemic model and degree distribution. Again, using Eq.~\eqref{eq:Y4rhok},  we have that the IPR  becomes finite indicating nonextensive localization in a finite number of vertices. Note that while a subextensive fraction of the network is active due to $\rho\sim N^{-1/(\gamma-1)}$ and, consequently, a diverging number of vertices is simultaneously infected as $N$ increases, the localization identified by the IPR is only in a finite number of nodes. This issue will be discussed further in the sequence of paper.

\begin{figure}
	\centering
	\includegraphics[width=0.97\linewidth]{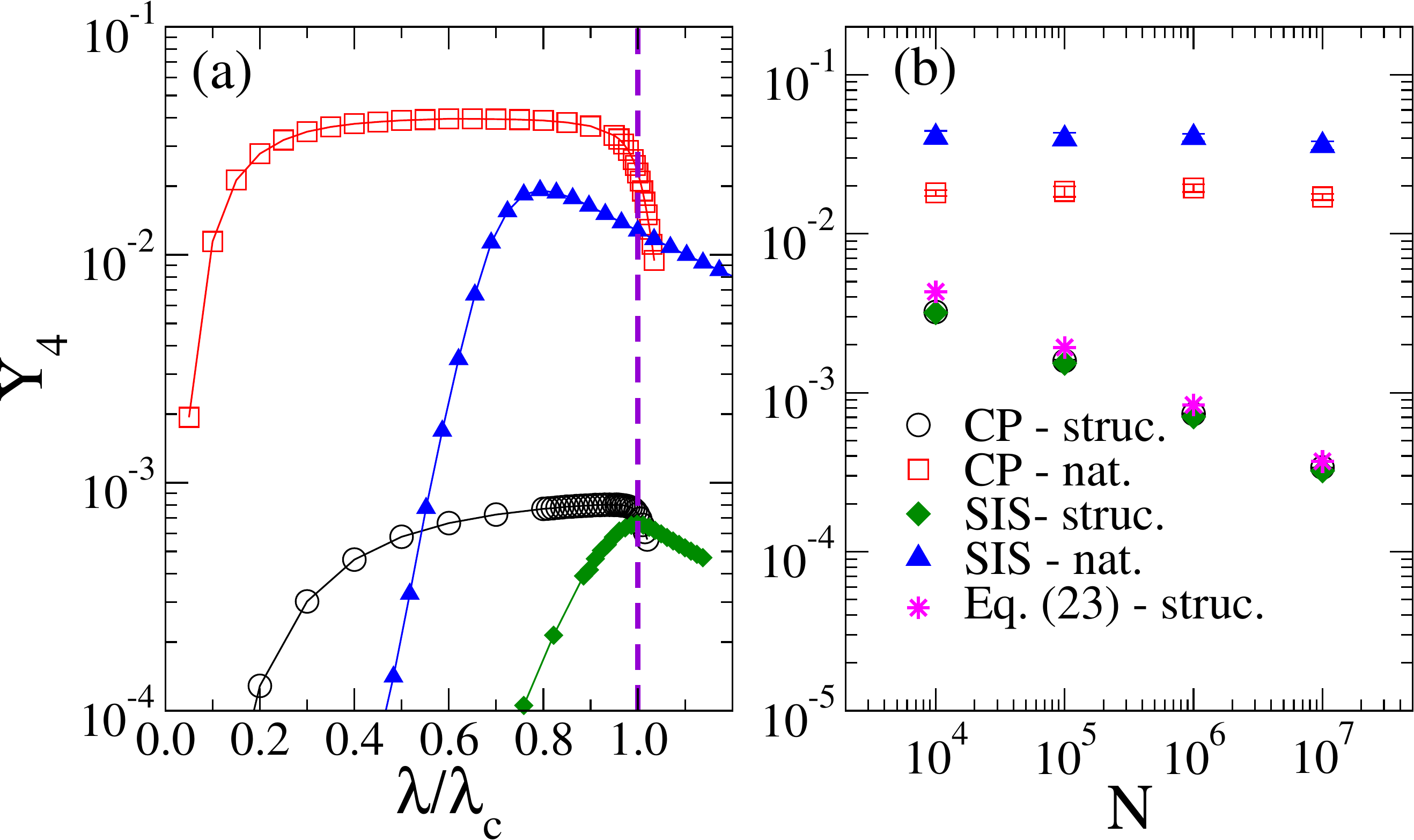}
	\caption{IPR analysis for SIS and CP dynamics on annealed networks with a power-law degree distribution of exponent $\gamma=2.3$ and two types of degree cutoff: natural ($\kmax\sim N^{1/(\gamma-1)}$) and structural ($\kmax\sim\sqrt{N}$). IPR dependence (a) with  infection rate  for fixed size $N=10^6$ and  (b)  with size for $\lambda=\lambda_\text{c}$ are shown. Dashed line represents $\lambda=\lambda_\text{c}$ in (a). IPR for structural cutoff computed with Eq.~\eqref{eq:Y4ann} is also shown in (b).}
	\label{fig:annealedSISandCPg230}
\end{figure}

We performed simulations of both SIS and CP with self-infection $f=\mu/N$ on annealed networks with power-law degree distributions using  natural and structural cutoffs. Figure~\ref{fig:annealedSISandCPg230} shows the IPR analyses for $\gamma=2.3$. Other values of $\gamma$ are shown  in comparisons with
quenched networks; see Sec.~\ref{subsec:quen}. The dependence of the IPR on infection rate is presented in Fig.~\ref{fig:annealedSISandCPg230}(a). The CP presents a faster drop to zero at $\lambda/\lambda_\text{c} \gtrsim 1$ in comparison with SIS model\footnote{The position of the susceptibility peak is slightly shifted from the mean-field prediction for SIS model with the natural cutoff. For CP this discrepancy does not happen.}. Notice that the abscissa is divided by $\lambda_\text{c}$ that goes to 0  and  $\mu$ for  SIS and CP, respectively. At the transition, Fig.~\ref{fig:annealedSISandCPg230}(b), the predictions of the mean-field theory are confirmed. In the case of a structural cutoff, both SIS and CP present the same IPR value approaching zero according to Eq.~\eqref{eq:Y4ann}.  For  $\gamma>5/2$, the scaling  given by Eq.~\eqref{eq:IPRstru} is confirmed but the prefactor is larger than the prediction of Eq.~\eqref{eq:Y4ann} as shown in Fig.~\ref{fig:dist_SIS350}.  For the natural cutoff, the IPR becomes finite confirming the localization predicted by the mean-field theory. Finally, the NAVCPD presents peaks at each vertex degree present in the network as will be shown in the comparison with the quenched case; see Sec.~\ref{subsec:quen}.

An interesting aspect calls our attention in this analysis. Activity in both CP and SIS dynamics in HMF or annealed approaches are localized at the transition according to the IPR analysis but an extensive, real endemic phase emerges above $\lambda_\text{c}$~\cite{Pastor-Satorras2001,Castellano2008}. So, we present examples where localization in the critical point does not discredit a mean-field theory for describing an endemic phase transition. A natural, but still unanswered question,  is whether such a reasoning applies to the QMF theory as well.

A central issue for dynamic process with localization is to understand  which subset of the network rules the epidemic prevalence and it was subject of intense studies~\cite{Goltsev2012, Lee2013, Boguna2013, Mata2015, Pastor-Satorras2018, Wei2020, Odor2014}. A misleading interpretation is that  the epidemic prevalence is given by the most localized region indicated by the IPR as we are going to disentangle from now on. In other words, we are going to show that a finite IPR, which is dominated by a few nodes, does not mean that the epidemic prevalence is also dominated by these few nodes. Let the fractional average of a  quantity $F(\phi)$  be defined  as
\begin{equation}
\zeta\lbrace F(\phi)\rbrace = \frac{\int_0^\phi F(\phi') \Pi(\phi') d\phi'}{\av{F(\phi)}},
\end{equation}
which yields the weight due to $\phi'<\phi$ in the average. For example, if $F=1$ we obtain the cumulative probability distribution. Since $\phi_i\propto \rho_i$ by definition, we have  $\zeta\lbrace \rho^n\rbrace = \zeta\lbrace \phi^n\rbrace$ and $\zeta\lbrace Y_4 \rbrace = \zeta\lbrace \phi^4 \rbrace$. The fractional average increases monotonically from 0 to 1 as $\phi$ varies from its minimal to maximal values. As an illustrative example, a fully delocalized state, in which $\Pi=\delta(\phi-1/\sqrt{N})$, has fractional averages given by a Heaviside function $\zeta=\Theta(\phi-1/\sqrt{N})$. 

\begin{figure}[th]
	\centering
	\includegraphics[width=0.988\linewidth]{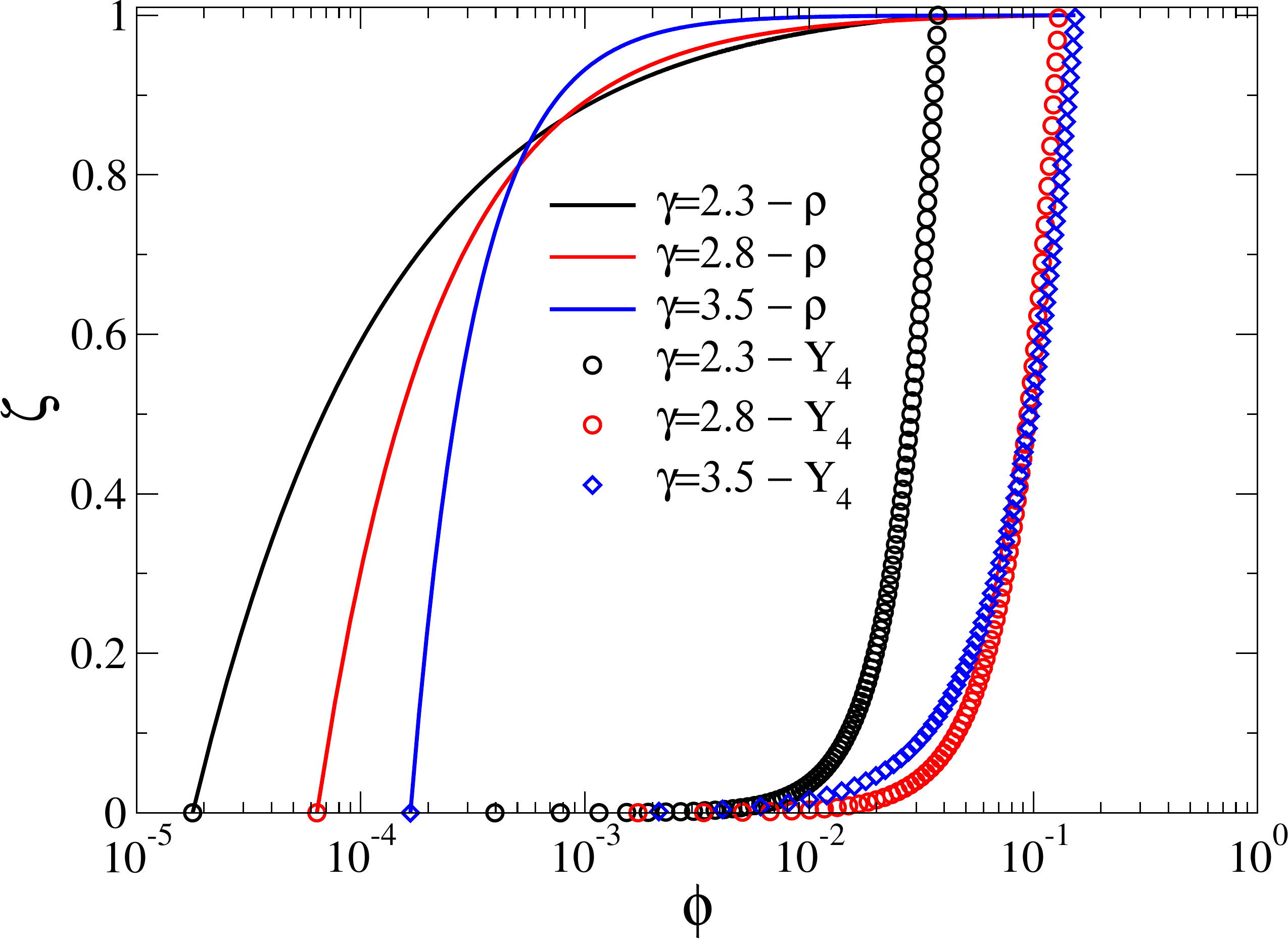}
	\caption{Fractional average analyses for prevalence $\rho$ (lines) and IPR $Y_4$ 	(symbols) for SIS using Eqs.~\eqref{eq:zetarhoann} and \eqref{eq:zetaY4ann} for size $N=10^7$, $\kc=2\sqrt{N}$, and different degree exponents.}
	\label{fig:weightN7ann}
\end{figure}

For HMF theory with a structural cutoff, the fractional average can be analytically computed using Eq.~\eqref{eq:phiann} to determine $\Pi(\phi)$ in terms of $P(k)$ resulting
\begin{equation}
\zeta\lbrace \rho \rbrace =
\frac{1-(\phi/\phi_0)^{2-\gamma}}{1-(\phi_\text{max}/\phi_0)^{2-\gamma}}
\label{eq:zetarhoann}
\end{equation}
and 
\begin{equation}
\zeta\lbrace Y_4 \rbrace = 
\frac{1-(\phi/\phi_0)^{5-\gamma}}{1-(\phi_\text{max}/\phi_0)^{5-\gamma}},
\label{eq:zetaY4ann}
\end{equation}
where $\phi_0$ and $\phi_\text{max}$ are given by Eq.~\eqref{eq:phiann}  with $k_i=k_0$ and $k_i=\kmax$, respectively. Figure~\ref{fig:weightN7ann} presents the fractional averages computed for HMF theory with different degree exponents and $\kc=2\sqrt{N}$ using Eqs.~\eqref{eq:zetarhoann} and \eqref{eq:zetaY4ann}.
The fractional averages for the IPRs become appreciable for $\phi$ near to $\phi_\text{max}$ when $\zeta\lbrace Y_4\rbrace\approx 1$, while for the prevalence it becomes appreciable just slightly above $\phi_0$ and is close to 1 when $\zeta\lbrace Y_4\rbrace$ is still departing from zero.

\begin{figure*}[!hbt]
	\centering
	\includegraphics[width=0.48\linewidth]{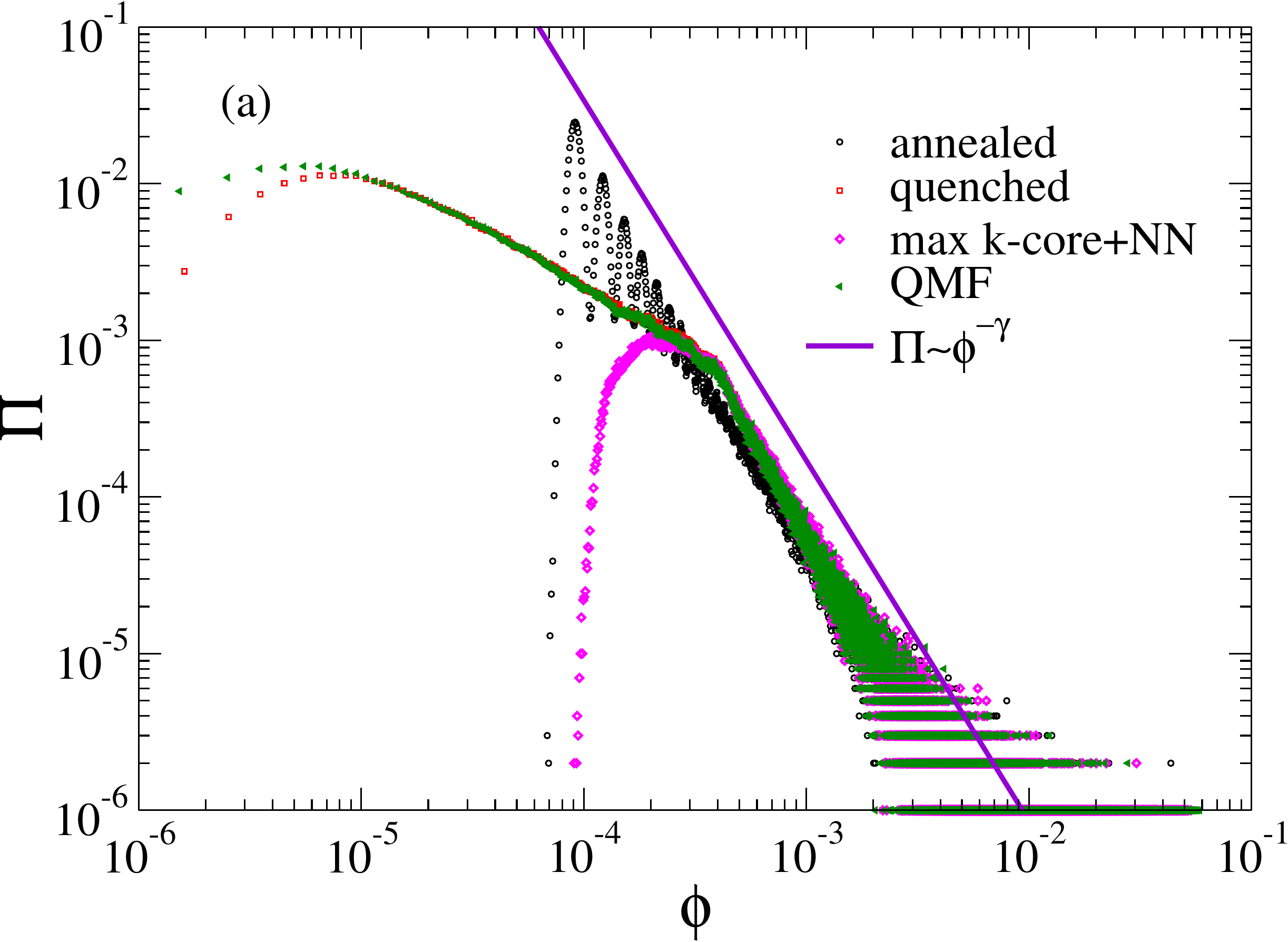}
	\includegraphics[width=0.48\linewidth]{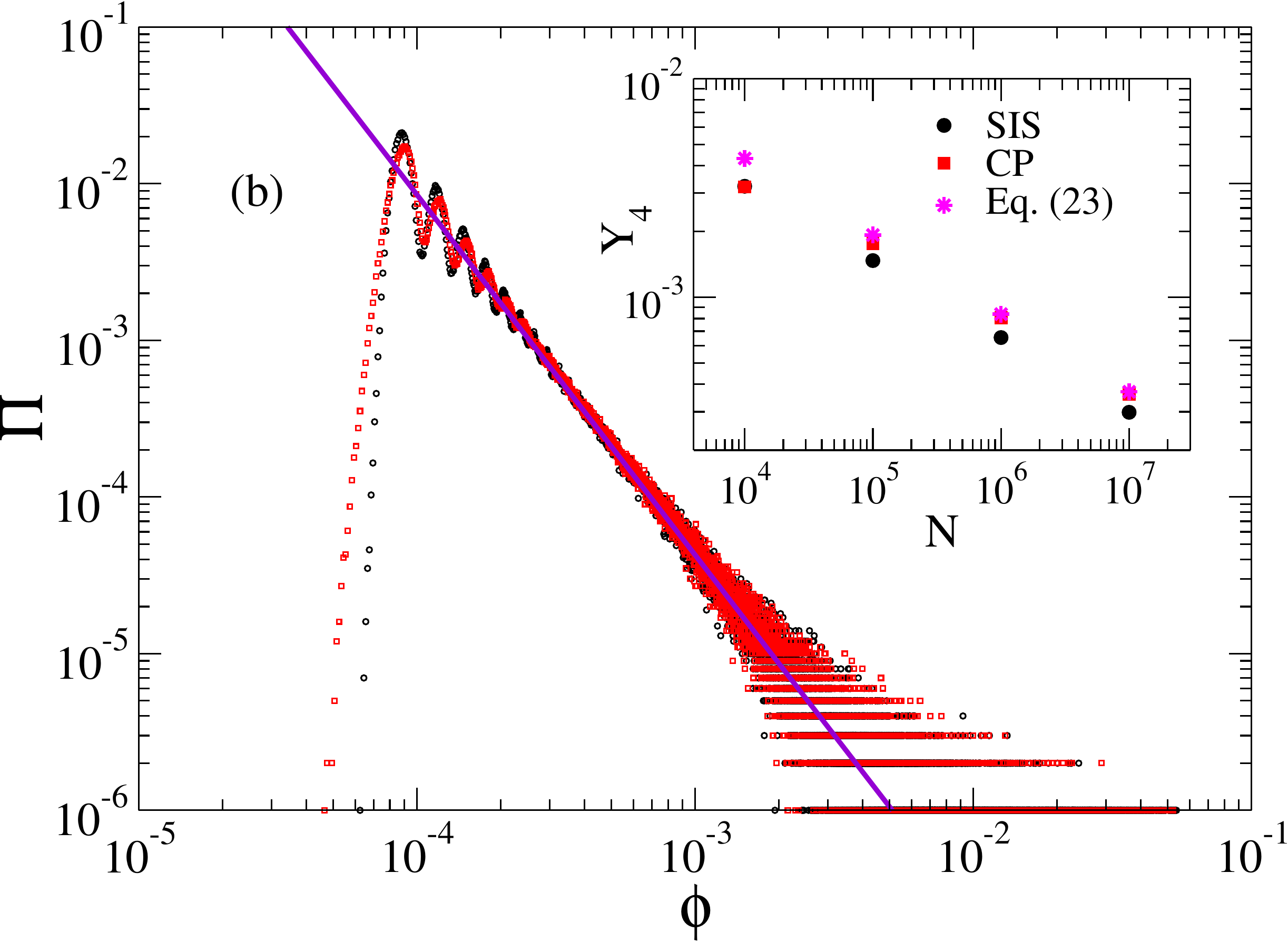}
	\caption{Analyses of the NAVCPD for simulations of (a) SIS and (b) CP models on a single sample of UCM or annealed networks of size $N=10^6$ and degree exponent $\gamma=2.3$. The curves were obtained at the epidemic threshold of the corresponding annealed ($\lambda_\text{c}^\text{CP}=1$ and $\lambda_\text{c}^\text{SIS}=0.0081$) and quenched ($\lambda_\text{c}^\text{CP}=1.12$ and $\lambda_\text{c}^\text{SIS}=0.0082$) networks, the latter for both QMF theory and simulations. The NAVCPD computed only  on the subgraph containing the vertices of the maximum $k$-core plus their nearest-neighbors (NN) is also shown (magenta symbols).   Inset: Finite-size scaling of the IPR for SIS  and CP models on UCM networks at the transition point. Both cases consider an upper cutoff $\kc=2\sqrt{N}$ for degree distribution.}
	\label{fig:dist_CP_SIS230}
\end{figure*} 	

The gap between fractional averages of prevalence and IPR increases with size implying that the epidemic prevalence is given by the delocalized component of the network. To show this, consider the value $\phi^*$ for which the fractional average corresponds to a fraction $(1-c)$ of total average, in which $c$ is small but finite (typically $c=0.1$), yielding an estimate for the range of the leading contributions to the averages. The corresponding ratio of $\phi^*$ for $\rho$ and $Y_4$ for $\gamma<5$ is
\begin{equation}
\frac{\phi^*{\lbrace{\rho}\rbrace}}{\phi^*\lbrace{Y_4}\rbrace}\simeq \dfrac{k_0}{\kmax}\dfrac{c^{-\frac{1}{(\gamma-2)}}}{(1-c)^{\frac{1}{5-\gamma}}}\ll 1.
\end{equation}
This result should be  understood as follows. While the IPR is dominated by a small, subextensive or nonextensive part of the system, relevant contributions for the order parameter come from the rest of the network. Moreover, as the network size increases the most active set contributes less for the overall activity and more for the IPR since $\kmax$ diverges with $N$.  The interpretation of localization near to a transition is made precise: While a small portion of the network can have extremely high activity that rules the IPR analysis, its contribution to the epidemic prevalence is negligible; the latter is given by the rest of the network that has lower level of activity but is much more numerous.

\subsection{Quenched networks}
\label{subsec:quen}

The simpler analytical toolbox based on HMF theory used for annealed networks is not available for the quenched case. Therefore, we use stochastic simulation~\cite{Cota2017} and numerical integration of the QMF equations~\cite{Silva2020}.  For the latter, we include the  self-activation term in  Eq.~\eqref{eq:QMF} to obtain
\begin{equation}
\frac{d\rho_i}{dt} = -\mu\rho_i+\lambda(1-\rho_i)\sum_{j=1}^{N}A_{ij}\rho_j+f(1-\rho_i),\label{eq:QMFsource}
\end{equation}
which was  solved numerically using a fourth order Runge-Kutta method. The QMF equation for CP can  be obtained from Eq.~\eqref{eq:QMFsource} replacing $A_{ij}$ by $A_{ij}/k_j$.

We used the uncorrelated configuration model (UCM)~\cite{Catanzaro2005} with a structural cutoff $\kc=2\sqrt{N}$ to generate quenched networks. Figure~\ref{fig:dist_CP_SIS230} shows the NAVCPD for simulations of the SIS and CP dynamics at their corresponding transition points running on UCM networks with  $\gamma=2.3$. Both models present very similar values of the IPR, consistent with the decays for annealed networks given by Eq.~\eqref{eq:Y4ann}; see inset of Fig.~\ref{fig:annealedSISandCPg230}(b). While the distribution for CP presents peaks corresponding to $\phi_i\propto k_i$ and matching almost exactly the annealed counterpart, the distribution for SIS matches almost perfectly the QMF theory, agreeing with  the annealed case only in the tails. The QMF theory for CP at the transition also presents the peaks $\phi_i\propto k_i$.  So, the IPR alone fails in distinguishing  differences in the localization of these dynamic processes. As discussed in Section~\ref{sec:models}, SIS and CP on quenched networks are characterized by different activation mechanisms~\cite{Ferreira2016a,Cota2018a} and differences in the localization are expected.

\begin{figure}[th]
	\centering
	\includegraphics[width=0.988\linewidth]{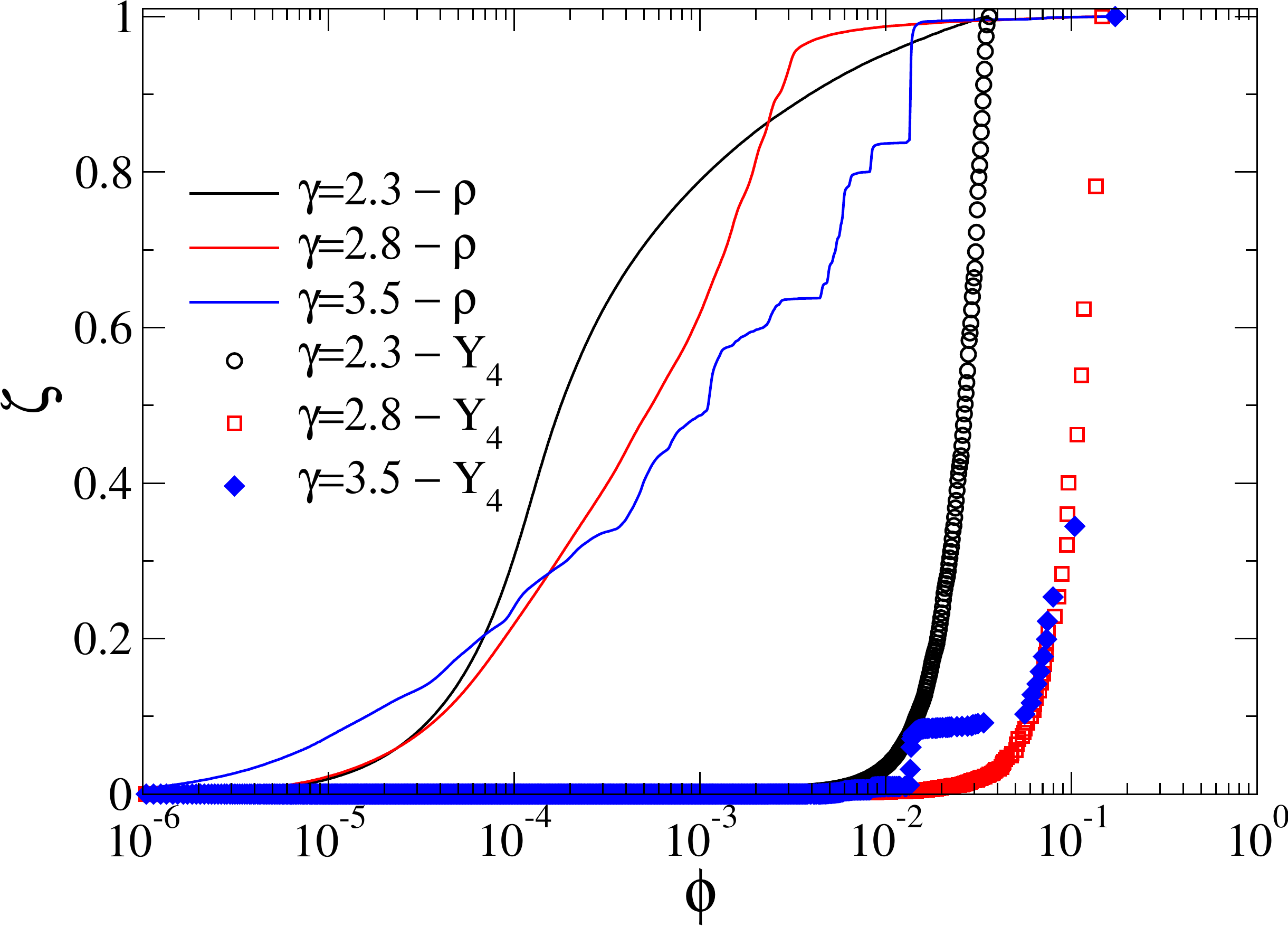}
	\caption{Fractional average analyses for prevalence $\rho$ (lines) and IPR $Y_4$ (symbols) for SIS at the transition point with self-infection $f=\mu/N$ run on single realizations of UCM networks of size $N=10^7$ and different degree exponents.}
	\label{fig:weightN7quen}
\end{figure}

The NAVCPD for critical SIS on quenched networks with $\gamma=2.3$ presents two asymptotic regimes:  It is approximately $\Pi\sim \phi^{-1}$ for low localization while the tail scales as does the degree distribution $\Pi\sim \phi^{-2.3}$. We numerically determined  that the crossover point between these regimes is very close to
\begin{equation}
\phi_*\approx\av{\phi}=\frac{1}{\sqrt{N}}\left(1-\frac{\av{\rho}}{\sqrt{\av{\rho^2}}}\right)~\sim N^{-1/2},
\end{equation} 
in which the equality can  be obtained from Eq.~\eqref{eq:phii}. This crossover is related to the nature of the activation of the SIS process for this range of $\gamma$, which is triggered in a densely connected component of the network identified by the maximum index of a $k$-core decomposition~\cite{Castellano2012}. The $k$-core decomposition is the removal of vertices and edges connected to them such that only vertices of degree $k\ge q$ remain in the network, where $q$ is the $k$-core index~\cite{Dorogovtsev2006a}. The maximum $k$-core corresponds to the last step before all vertices are removed. Figure~\ref{fig:dist_CP_SIS230}(a) shows the NAVCPD computed for SIS only on a subgraph containing the vertices belonging to the maximum $k$-core plus their nearest neighbors. The distribution in this subgraph matches very well the overall distribution for $\phi>\phi_*$, dropping quickly for $\phi<\phi_*$, which corroborates our assertion. Note that the maximal $k$-core itself corresponds only to the narrow end of the tail. This subset corresponds to a vanishing fraction of the network as $N \rightarrow \infty$~\cite{Cota2018a} that dominates the contributions for IPR; see discussion on Fig.~\ref{fig:weightN7quen}.

Some important consequences come from Fig.~\ref{fig:dist_CP_SIS230}  and  analyses of CP for other values of the degree exponents. A first one is the corroboration that CP on quenched networks is a HMF-like process~\cite{Ferreira2011,Mata2014}, an issue intensely debated in the 2010s~\cite{Castellano2006, Hong2007, Ha2007, Castellano2007, Castellano2008}, due to the almost perfect match of NAVCPD for simulations on quenched and annealed networks. For SIS, a remarkable agreement between NAVCPD for QMF  theory and simulations occurs showing that the QMF is able to capture almost perfectly the localization pattern observed in simulations for the investigated range size (up to $N= 10^7$), even though the critical exponents of the prevalence are different with $\beta> \beta_\text{QMF} = 1$~\cite{Mountford2013,Silva2019}.  Notice that a similar situation happens for CP and SIS on annealed networks as discussed in Sec.~\ref{subsec:ann}. Obviously, one cannot discard the possibility of a different scenario  for much larger, computationally inaccessible sizes. Finally, the crossover in the NAVCPD of SIS also indicates that the HMF behavior holds for an extended maximum $k$-core that encloses its nearest-neighbors.

\begin{figure}[tbh]
	\centering
	\includegraphics[width=0.988\linewidth]{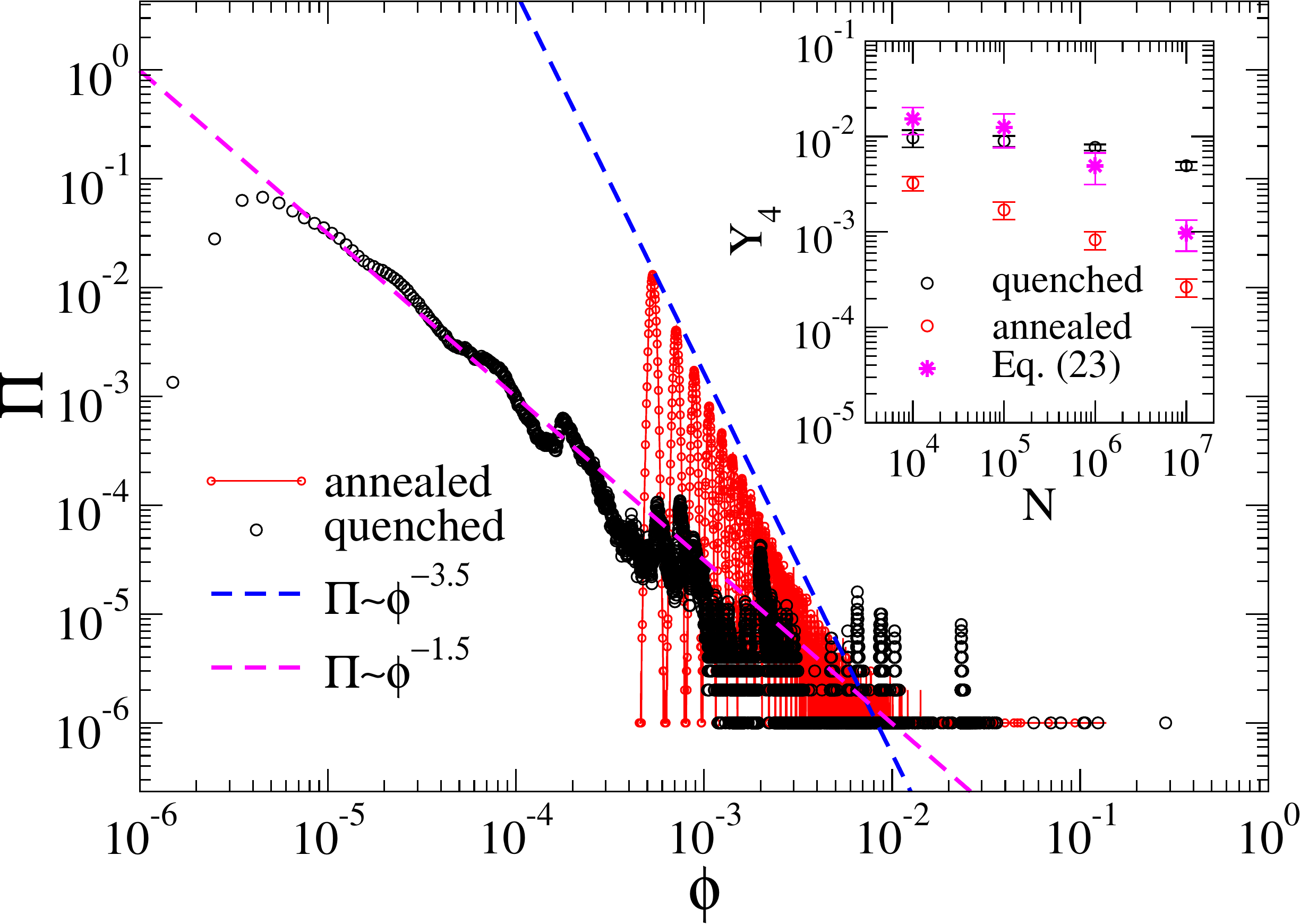}
	\caption{Analyses of the NAVCPD for SIS model on UCM and annealed networks  	(single realization of size $N=10^6$) with degree exponent $\gamma=3.5$ using $\kc=2\sqrt{N}$ which leads to a  highly  fluctuating cutoff $\kmax\sim N^{1/(\gamma-1)}$. The inset shows the finite-size scaling for the IPR estimated at the maximal value of the dynamical susceptibility, averaged over 10 and 50 independent network realizations of quenched and annealed networks, respectively.  IPR computed using Eq.~\eqref{eq:Y4ann} is also shown.
	}
	\label{fig:dist_SIS350}
\end{figure}

\begin{figure*}[hbt]
	\centering
	\includegraphics[width=0.46\linewidth]{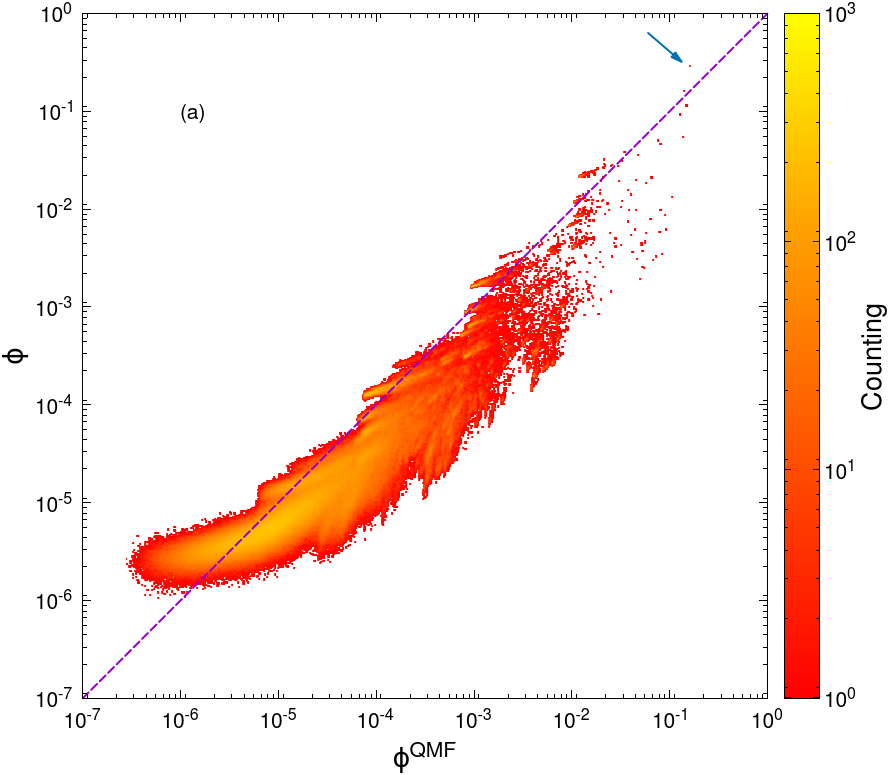}~~
	\includegraphics[width=0.46\linewidth]{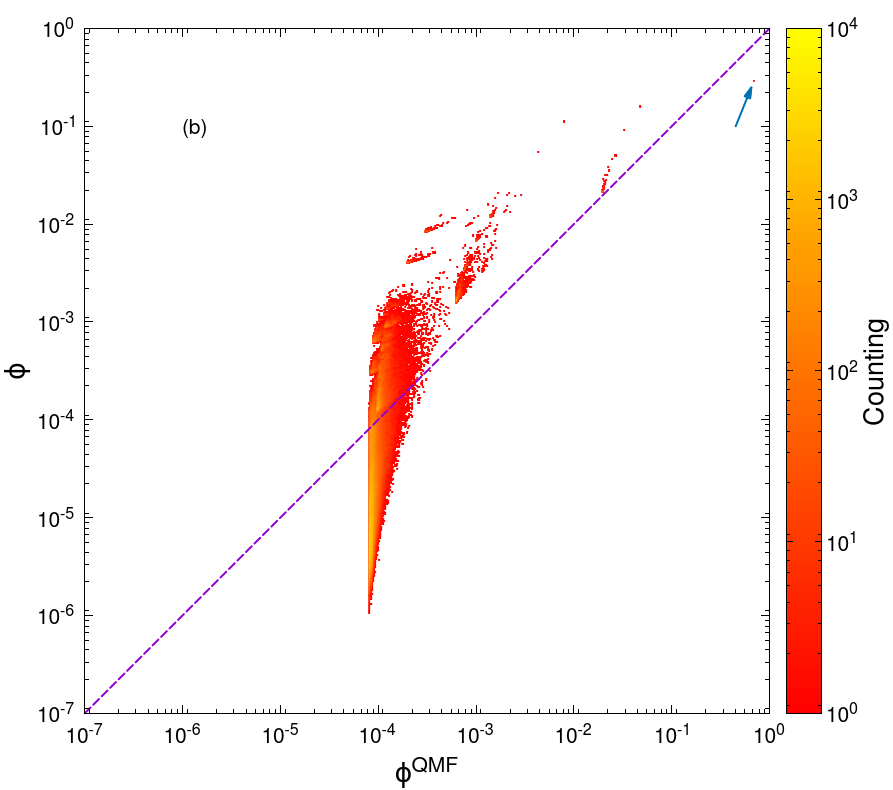}
	\caption{Binned scatter plots (grid with $500\times 500$ bins in a logarithm scale) for NAV components at the transition point obtained in stochastic simulation (ordinate) and QMF theory (abscissa) on UCM networks  $\kc=2\sqrt{N}$) of degree exponent $\gamma=3.5$. Color bars represent the amount of points within a bin. Simulations for $\lambda=\lambda_\text{c}=0.0775$, given by the dynamical susceptibility, are compared with (a) integration of the QMF equations for this same value of $\lambda$ and (b) and with the critical QMF theory at $\lambda=\lambda_\text{c}^\text{QMF}=1/\Lambda_1=0.0225$. Network size is 	$N=10^6$. Dashed line is the identity $\phi=\phi^\textrm{QMF}$.}
	\label{fig:scatter}
\end{figure*}

For $\gamma>5/2$ striking divergences between SIS dynamics on annealed and quenched networks emerge~\cite{Mountford2013, Ferreira2012, Castellano2012}, the more evident for the larger $\gamma$. Figure~\ref{fig:dist_SIS350} shows the NAVCPD analyses of the critical SIS for $\gamma=3.5$ on both quenched and annealed networks with a same degree distribution. For annealed networks, the behavior is similar to the case $\gamma=2.3$ with peaks proportionally related to the vertex degree. The corresponding IPR decays according to Eq.~\eqref{eq:Y4ann} as $Y_4\sim N^{-0.4}$, but the prefactor is not accurate as it was for $\gamma=2.3$; see Fig.~\ref{fig:dist_CP_SIS230}(b). Simulations on annealed networks are less localized than the HMF theory?s prediction due to the fluctuations absent in theory which contribute to spread the activity reducing, therefore, the localization. 
In the case of quenched UCM networks, the IPR seems to converge to a finite value but it has not achieved its asymptotic value in the investigated size range with $Y_4\approx 0.01$ for $N=10^7$. The NAVCPD for quenched networks differs substantially from their annealed counterparts. No correlation between NAV  and degree distribution is evident. Also, it does not quantitatively match with QMF theory  either, but it seems to capture the trend; see discussion on Fig.~\ref{fig:scatter}. The NAVCPD for the quenched network  presents very heavy tails, decaying approximately as $\Pi\sim \phi^{-\eta}$ with $\eta\lesssim 2$ (the actual $\eta$ value is beyond our scope), being therefore ruled by outliers that determine the IPR value. The case $\gamma=2.8$, which is not shown for brevity,  is similar to $\gamma=3.5$.

The fractional averages of prevalence and IPR for  SIS at $\lambda=\lambda_\text{c}$ are presented in Fig.~\ref{fig:weightN7quen} for single realizations of  UCM networks with $N=10^7$ vertices and three  values of the degree exponent. As in the annealed case, we have that vertices of lowest $\phi$ values contribute significantly for prevalence while the IPR is determined by the most active ones; the latter contribute negligibly for the prevalence value. Curves for $\gamma=2.3$ are smooth and qualitatively similar to the annealed case, with significant differences only for prevalence at very low $\phi$. In turn, curves for larger values of $\gamma$ contrast with the annealed case,  being more evident for larger $\gamma$, in which the fractional IPRs vary abruptly for large $\phi$. In Fig.~\ref{fig:weightN7quen}, for example, the vertex of highest $\phi$ represents 22\% and 66\% of the total IPR for networks with $\gamma=2.8$ and 3.5, respectively. Curve $\zeta\lbrace\rho\rbrace$ for $\gamma=3.5$ presents several  jumps and plateaus. The same happens for the corresponding IPR, which can be seen in a (not shown) double-logarithm scale. This is an explicit observation of localized activity that is known to drive the epidemic activation for SIS at UCM networks with $\gamma>3$~\cite{Chatterjee2009,Boguna2013,Ferreira2016a}.

Important insights and elucidations are extracted from Figs.~\ref{fig:dist_CP_SIS230}, \ref{fig:dist_SIS350}, and \ref{fig:weightN7quen}, helping to rationalize  whether metastability of weakly interacting subgraphs leads or not to an actual endemic phase~\cite{Goltsev2012,Lee2013,Boguna2013, Mata2015, Pastor-Satorras2018, Wei2020, St-Onge2020}. The SIS dynamics on quenched networks at the epidemic threshold is indeed localized for all values of $\gamma$ (subextensively for $\gamma<5/2$ and on a finite set for $\gamma>5/2$) as does the PVE of the adjacency matrix~\cite{Pastor-Satorras2016}, in the sense that a vanishing minority of vertices presents extremely high epidemic activity, ruling the IPR analysis. However, the corresponding epidemic process can still be undergoing a phase transition, in which extensive components  are participating in the processes.

Figure~\ref{fig:scatter}(a) presents log-binned scatter plots for NAV components obtained in SIS simulations ($\boldsymbol{\phi}$) and integration of the QMF equations  ($\boldsymbol{\phi}^\text{QMF}$)  for an infection rate $\lambda=\lambda_\text{c}$ determined by the maximum of $\chi$. In contrast with the case $\gamma=2.3$, for which an almost perfect match is observed (see Fig.~\ref{fig:dist_CP_SIS230}), the scatter plots for $\gamma=3.5$ present a very disperse distribution of points. The  integration captures the trend of the simulation presenting, however, most vertices with NAV components higher than in stochastic simulations, which is indicated by most points below the identity line. This is due to the QMF equation being actually supercritical since $\lambda_\text{c}>\lambda_\text{c}^\text{QMF}$. In Fig.~\ref{fig:scatter}b), we compare the situation in which both QMF theory and stochastic simulations  are in their corresponding epidemic thresholds ($\lambda=\lambda_\text{c}$ for simulations and $\lambda = \lambda_\text{c}^\text{QMF} = 1/\Lambda_1$ for QMF theory). Here, the converse of Fig.~\ref{fig:scatter}(a) is observed: Activity in QMF theory is slightly correlated with stochastic simulations only for the few and most active vertices whereas the vast majority of the network presents no correlation. 

In order to illustrate the distinct levels of localization in these approaches, consider the single vertex contributions to the IPR  in the network sample shown in Fig.~\ref{fig:scatter}. The vertex indicated by arrows is the one of higher activity in both simulations and QMF theory. However, the gap between the  first, $\phi_1$, and second, $\phi_2$, most active vertices is much larger in critical QMF than in simulations, given by $\phi_1/\phi_2\approx 15$ and $1.8$, respectively. In both cases, the IPR is led by the most localized vertex but much more in the QMF theory. Indeed, this single vertex contributes with $(\phi_\textrm{1}^\text{QMF})^4 = 0.2380$ to the IPR $Y_4(\boldsymbol{\phi}^\text{QMF})=0.2384$ for critical QMF theory, and with $\phi_1^4 =0.0074$  of $Y_4(\boldsymbol{\phi})=0.0087$ for stochastic simulations. For supercritical QMF analyzed in Fig.~\ref{fig:scatter}a), we have $(\phi^\text{QMF}_1)^4=0.0007$ of $Y_4(\boldsymbol{\phi}^\text{QMF}) = 0.0029$. Note that the IPR of critical QMF is very close to the value $Y_4=1/4$ of a star graph; see Section~\ref{subsec:star}. 

From the perspective of theory accuracy, these results show that QMF theory should be used with extreme caution to determine the most active vertices in the network since  it may deviate substantially from  the actual picture as shown in Fig.~\ref{fig:scatter}a). An alternative is to use pair QMF theory~\cite{Mata2013,Silva2020} which reckons dynamical correlations considering pairwise equations. Indeed, a pair QMF theory  was recently used to identify the most central spreaders that should be immunized to efficiently stop  a SIS dynamics~\cite{Matamalas2018}. However, even pair QMF theory has limitations for SIS dynamics on networks with large $\gamma$~\cite{Silva2019} and should also be used with caution.

\section{Conclusions}
\label{sec:conclu}

Effects of localization introduced by structural disorder play a major rule on equilibrium and nonequilibrium statistical mechanics~\cite{Vojta2006} and have recently been applied to dynamical processes on networks~\cite{Goltsev2012,Moretti2013,Cota2018,Buono2013,Hebert-Dufresne2019}, in which disorder is intrinsic to the heterogeneous connectivity structure observed in real and synthetic networks. The activity localization  in dynamical processes has been widely, not exclusively, investigated in terms of spectral properties of Jacobian matrices that emerge in the linearization of mean-field equations near to the transition~\cite{Mieghem2012, Goltsev2012, Pastor-Satorras2016, Castellano2018, Martin2014, Silva2019}. In the present work, we develop a localization analysis applicable to any type of dynamical process for which  activity can be gauged in terms of a local order parameter which, in present work, was considered at a vertex level but lower resolution motifs are eligible as well. The method is generic and can be applied to both mean-field and stochastic simulations whether the system is near to a transition or not.

We applied the methodology to the SIS~\cite{PastorSatorras2015} and CP~\cite{Castellano2006} models, two conceptually similar dynamical processes with very distinct activation mechanisms, to a wide spectrum of random networks which were handled both analytically (mean-field theories) or using simulation tools. We were able to disentangle some misleading interpretations of localization in epidemic models. One conclusion is that the IPR, a commonly used metric to quantify localization in an overall level, blurs the structured localization patterns present in complex and even relatively simple networks.

Handling the HMF theories for SIS and CP on networks with power-law degree distributions, we have highlighted several elucidative outcomes. We show that SIS and CP present exactly the same localization  patterns at the epidemic threshold  for any value of the degree exponent $\gamma$, despite these models being characterized by different critical exponents for the epidemic prevalence when $\gamma<4$~\cite{Pastor-Satorras2001,Castellano2006}. Also, we show that both CP and SIS dynamics at their respective thresholds are localized in a subextensive subset (increasing sublinearly with network size) for $\gamma<5$ and concomitantly give rise to an endemic (delocalized) phase immediately above the threshold. Finally, defining the concept of fractional averages we show that even in states which are localized according to the IPR analysis, the order parameter is determined by the vast majority of vertices in which the activity is smaller in contrast with the IPR which is ruled by a fewer (subextensive), small fraction of the network. Even seeming unsound at a first glance, the interpretation is quite simple. A few vertices are active most of the time while others are  active only eventually due to stochastic interactions with the former. However, the latter being infinitely larger in number will lead any intensive quantity such as the order parameter. These finds are foundational for the correct interpretation of activation mechanisms of  endemic phases.

We also analyzed localization of both dynamics on quenched networks, which are in general not accurately described by mean-field approaches and were handled using stochastic simulations. Firstly, we corroborate the HMF-like nature of the CP on quenched networks, an issue widely debated formerly~\cite{Castellano2006, Castellano2007, Castellano2008, Ha2007, Hong2007} in terms of mean-field exponents, showing that annealed and quenched versions of a same degree distribution have the same localization patterns depending only on the vertex degree. We stress again that the localization pattern alone cannot give the ultimate response and it complements other evidence reported elsewhere~\cite{Ferreira2011,Mata2014}.

The  results for SIS on quenched networks are richer and more complex. For small $\gamma<5/2$, the localization patterns at the threshold  match almost perfectly the QMF theory, agreeing with the annealed case only for  a subset essentially composed by the  vertices of the maximum $k$-core plus its nearest neighbors, being still a subextensive subgraph. Again, one knows that critical exponent of the epidemic prevalence of the QMF theory  $\beta_\textrm{QMF}=1$ is not the correct one since $\beta>1$ for any value of $\gamma>2$~\cite{Mountford2013}. Despite this,  our numerical results suggest that  QMF is still able to capture accurately the localization structure. For $\gamma>5/2$, the distribution of activity localization is more complex presenting outliers and heavy tailed distributions that lead to epidemic localization in  a finite subset, as predicted by the QMF theory without however an accurate quantitative match with the simulations. Indeed, comparing critical behavior of both simulations and QMF theory, correlation is found only for a tiny part of the vertices in which activity is highly localized.

To sum up, our results show that the localization patterns are much more complex and revealing than simple metrics such as the IPR can tell. Using spectral properties of, for example, adjacency or non-backtracking matrices, derived within mean-field frameworks, can blur important features of the dynamical process and also lead to misleading  or incomplete conclusions on the physical mechanisms involved in. The methodology studied here can be extended to any class of networks and dynamical processes. Therefore, we expect many forthcoming analyses of localization of other dynamical processes, such as opinion dynamics~\cite{Castellano2009} and synchronization~\cite{Rodrigues2016}, as well as other network topologies  such as multilayer~\cite{DeArruda2017a} and temporal~\cite{Holme2015} networks.

\begin{acknowledgments}
This work was partially supported by the Brazilian agencies \textit{Coordenação
	de Aperfeiçoamento de Pessoal de Nível Superior} - CAPES (Grant no.
88887.507046/2020-00), \textit{Conselho Nacional de Desenvolvimento Científico e
	Tecnológico}- CNPq (Grants no. 430768/2018-4 and 311183/2019-0) and
\textit{Fundação de Amparo à Pesquisa do Estado de Minas Gerais} - FAPEMIG
(Grant no. APQ-02393-18). This study was financed in part by the
\textit{Coordenação de Aperfeiçoamento de Pessoal de Nível Superior} (CAPES) -
Brasil  - Finance Code 001.
\end{acknowledgments}


%

\end{document}